\documentclass[aps,prb,twocolumn,superscriptaddress,floatfix]{revtex4}
\usepackage{bm}
\usepackage{amsmath}
\usepackage{amssymb}
\usepackage{graphicx}
 \usepackage{color}
\usepackage{soul,xcolor}
\setstcolor{red}
\usepackage{hyperref}

\usepackage{xcolor}\definecolor{ao}{rgb}{0.0,0.0,1.0}
\definecolor{br}{rgb}{1.0, 0.22, 0.0}

\usepackage{epsfig}

\input{epsf}

\newcommand{\kB}{k_{\text{B}}}
\newcommand{\deriv}[2]{\frac{\text{d}{#1}}{\text{d}{#2}}}

\begin{document}

\title{Thermoelectricity near  Anderson localization transitions}

\author{Kaoru Yamamoto}
\affiliation{Department of Physics, The University of Tokyo, Kashiwanoha 5-1-5, Kashiwa, Chiba 277-8574, Japan}
\email{kaoru3@iis.u-tokyo.ac.jp}

\author{Amnon Aharony}
\affiliation{Raymond and Beverly Sackler School of Physics and Astronomy, Tel Aviv University, Tel Aviv 69978, Israel}
\affiliation{Physics Department, Ben Gurion University, Beer Sheva 84105, Israel}
\email{aaharony@bgu.ac.il}

\author{Ora Entin-Wohlman}
\affiliation{Raymond and Beverly Sackler School of Physics and Astronomy, Tel Aviv University, Tel Aviv 69978, Israel}
\affiliation{Physics Department, Ben Gurion University, Beer Sheva 84105, Israel}

\author{Naomichi Hatano}
\affiliation{Institute of Industrial Science,  The University of Tokyo, Kashiwanoha 5-1-5, Kashiwa, Chiba 277-8574, Japan}

\date{\today}

\begin{abstract}

The electronic thermoelectric coefficients are analyzed in the vicinity of one and two Anderson localization thresholds in three dimensions. For a single mobility edge, we correct and extend previous studies and find universal approximants which allow us to deduce the critical exponent for the zero-temperature conductivity from thermoelectric measurements. In particular, we find that at nonzero low temperatures the Seebeck coefficient and the thermoelectric efficiency can be very large on the ``insulating" side, for chemical potentials  below the (zero-temperature) localization threshold. Corrections to the leading power-law singularity in the zero-temperature conductivity are shown to introduce nonuniversal temperature-dependent corrections to the otherwise universal functions which describe the Seebeck coefficient, the figure of merit and the Wiedemann-Franz ratio.  Next, the thermoelectric coefficients are shown to have interesting dependences on the system size. While the Seebeck coefficient decreases with decreasing size,  the figure of merit, first decreases but then increases, while  the Wiedemann-Franz ratio first increases but then decreases as the size decreases. Small (but finite) samples may thus have larger thermoelectric efficiencies. In the last part we study thermoelectricity  in systems with a pair of localization edges, the ubiquitous situation in random systems near the centers of electronic energy bands. As the disorder increases, the two thresholds approach each other, and then the Seebeck coefficient and the figure of merit increase significantly, as expected from the general arguments of Mahan and Sofo  [J. D. Mahan and J. O. Sofo, Proc. Natl. Acad. Sci. U.S.A. {\bf 93}, 7436 (1996)] for a narrow  energy range of the zero-temperature metallic behavior.

\end{abstract}

\maketitle

\section{Introduction}
\label{Intro}

Thermoelectric coolers are based on the Peltier effect, which predicts the appearance of a heat current when an electric current is passed through a material.\cite{goldsmith} Alternatively, the Seebeck effect generates an electric current from a temperature gradient. For a recent review, see Ref.~\onlinecite{benenti-rev}.
When a thermodynamic system is connected to two electronic reservoirs, the voltage $V$ (equal to the electrochemical potential difference divided by the  electron charge $e$) and the temperature difference $\Delta T=T_H-T_C$ between the (hot and cold, $T_H>T_C$) reservoirs drive a charge current $j_e$ and a heat current $j_h$ between  them.
In the linear-response regime, the charge and heat currents are related to the thermodynamic driving  forces (i.e.,  the voltage and the temperature difference) via the Onsager matrix, which is symmetric for time-reversal invariant systems,\cite{onsager,callen}
\begin{align}
\begin{pmatrix}j_e \\ j_h\end{pmatrix} = \begin{pmatrix}L_{11} & L_{12} \\ L_{12} & L_{22} \end{pmatrix}\begin{pmatrix}V \\ \Delta T/T\end{pmatrix},
\label{Eq1}
\end{align}
where $T$ is the average temperature of the reservoirs (and of the system).
The commonly observed transport properties, namely the electrical conductivity $\sigma$, the thermopower (or Seebeck coefficient) $S$, and the electronic heat conductivity (in the absence of charge current) $\kappa$,\cite{com} are related to the elements of the Onsager   matrix as follows:
\begin{align}
\sigma&=L_{11},\nonumber\\
S&=L_{12}/(TL_{11}),\nonumber\\
\kappa&=(L_{11}L_{22}-L_{12}^2)/(TL_{11}).
\label{SSS}
\end{align}
 In the presence of time-reversal symmetry, the  Peltier coefficient $\Pi$ and the Seebeck coefficient $S$ are related via the Onsager relation,
\begin{align}
\Pi=TS=L_{12}/L_{11}.
\label{SSP}
\end{align}

The  thermodynamic efficiency $\eta$ of a heat engine which transforms heat $Q$ (flowing out of the hot reservoir) into electric work $W$ is defined as the ratio $\eta=W/Q=P/j_h$, where $P=dW/dt=j_eV$ is the electric power.
This efficiency is bounded by the Carnot efficiency, $\eta\le\eta^{}_C=\Delta T/T_H=1-T_C/T_H$.\cite{goldsmith,benenti-rev,yamamoto2015} Much of the research on thermoelectricity is aimed at finding devices with a high efficiency. For systems with time-reversal symmetry, the maximum efficiency is given by
\begin{align}
\eta^{}_{\rm max}=\eta_C\frac{\sqrt{ZT+1}-1}{\sqrt{ZT+1}+1},
\label{eta}
\end{align}
where $ZT$ is the dimensionless figure of merit,\cite{goldsmith,benenti-rev}
\begin{align}
ZT=\frac{\sigma S^2}{\kappa}T=\frac{L_{12}^2}{L_{11}L_{22}-L_{12}^2}.
\label{ZT}
\end{align}

Clearly, $\eta^{}_{\rm max}$ approaches the Carnot efficiency $\eta^{}_C$ when $ZT$ approaches $\infty$. Unfortunately, high values of $ZT$ are difficult to achieve, and it is said that a good thermoelectric device should have $ZT>3$.\cite{benenti-rev}
In  the definition of $ZT$,  the central part of Eq.~(\ref{ZT}),  the heat conductivity $\kappa$  consists of {\it both} the electronic heat conductivity and the phononic heat conductivity. In the present paper we mostly concentrate on the electronic heat conductivity, and assume sufficiently low temperatures, so that we can ignore the phononic heat conductivity. However, see  the discussion in the last section.

Another quantity of interest is the  Wiedemann-Franz ratio,
\begin{align}
{\cal L}=\frac{\kappa}{\sigma T}\equiv \frac{S^2}{ZT}.
\label{WF}
\end{align}
For  the electronic charge and heat conductivities in metals, the Sommerfeld expansion yields the universal Lorenz value \cite{ashcroft}  ${\cal L}_0=(\pi^2/3)\kB^2/e^2$. However, as we discuss below, in many cases one encounters smaller values of ${\cal L}$, which may imply larger values of $ZT$.

Ignoring inelastic phononic effects, the electronic {\bf linear-response coefficients} $L_{ij}$ are obtained with the Chester-Thellung-Kubo-Greenwood (CTKG) formulation \cite{kubo1957, greenwood1958, chester1961} as
\begin{align}
L_{11}&= \int_{-\infty}^{\infty} dE \sigma^{}_0(E) F(E), \notag\\
L_{12} &=\frac{1}{|e|}\int_{-\infty}^{\infty} dE (E-\mu)\sigma^{}_0(E) F(E), \notag\\
L_{22} &= \frac{1}{e^2}\int_{-\infty}^{\infty} dE (E-\mu)^2 \sigma^{}_0(E) F(E),
\label{LL}
\end{align}
where
\begin{align}
F(E) = -\deriv{f(E)}{E} \equiv \frac{e^\epsilon}{\kB T(1+e^\epsilon)^2}= \frac{1}{4\kB T\cosh^2(\epsilon/2)}
\end{align}
is the negative derivative of the Fermi function  with respect to the energy $E$, $f(E)=(1+e^{\epsilon})^{-1}$, with $\epsilon=(E-\mu)/k_BT$.
Here $\kB$ is the Boltzmann constant, while $T$ and $\mu$ are the  (common) temperature and the electrochemical potential of the system, respectively. Note that $\mu$ depends on temperature, coinciding with the Fermi energy $E_F$ only at $T=0$.\cite{villagonzalo1999}  Our results are expressed in terms of $\mu$, and thereby this subtle point is circumvented.
In Eq.~(\ref{LL}), $\sigma^{}_0(E)$ is the  zero-temperature conductivity of the system, which contains both the electronic density of states and the Landauer transmission through the system.

\begin{figure}
    \includegraphics[width=6cm]{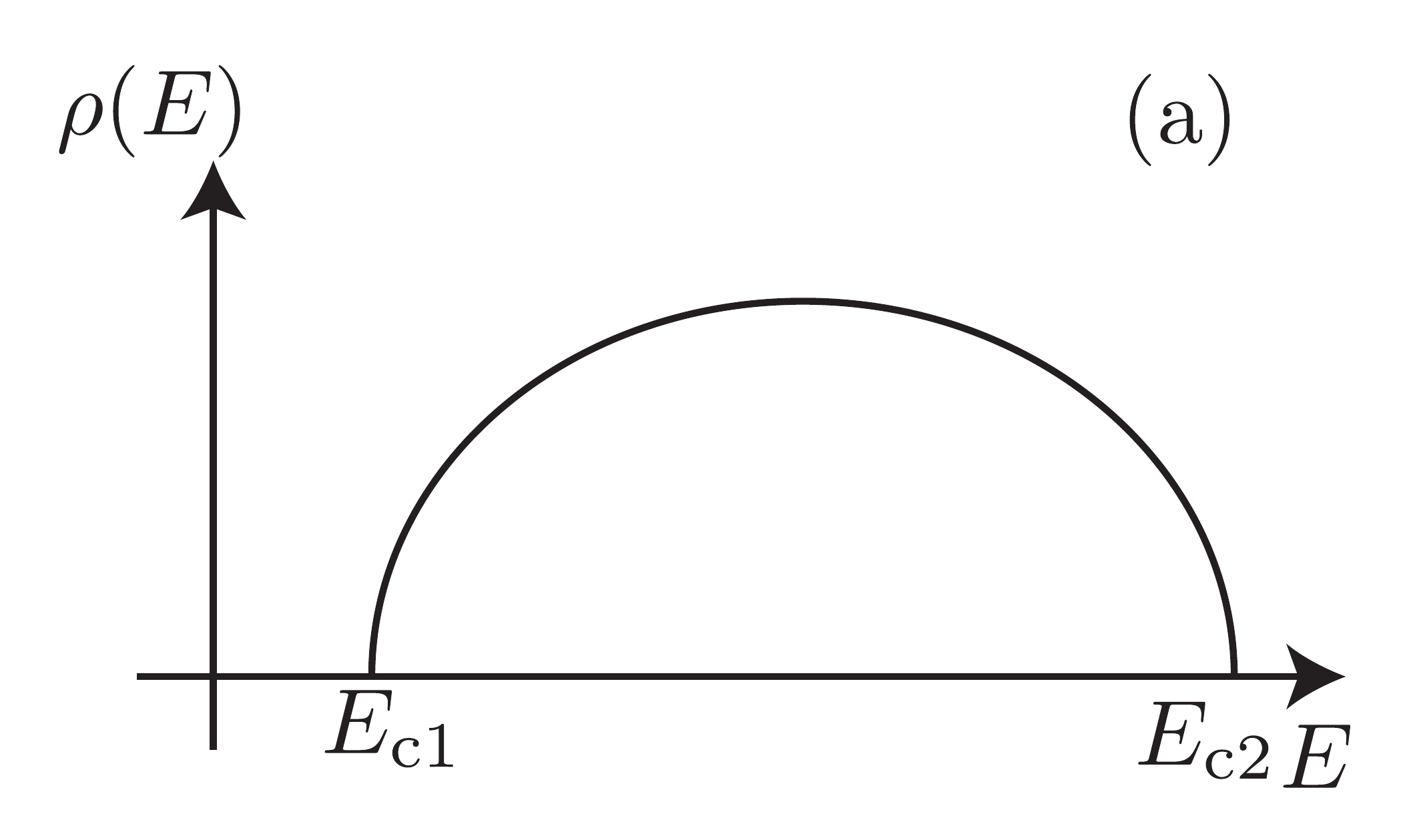}\\
    \includegraphics[width=6cm]{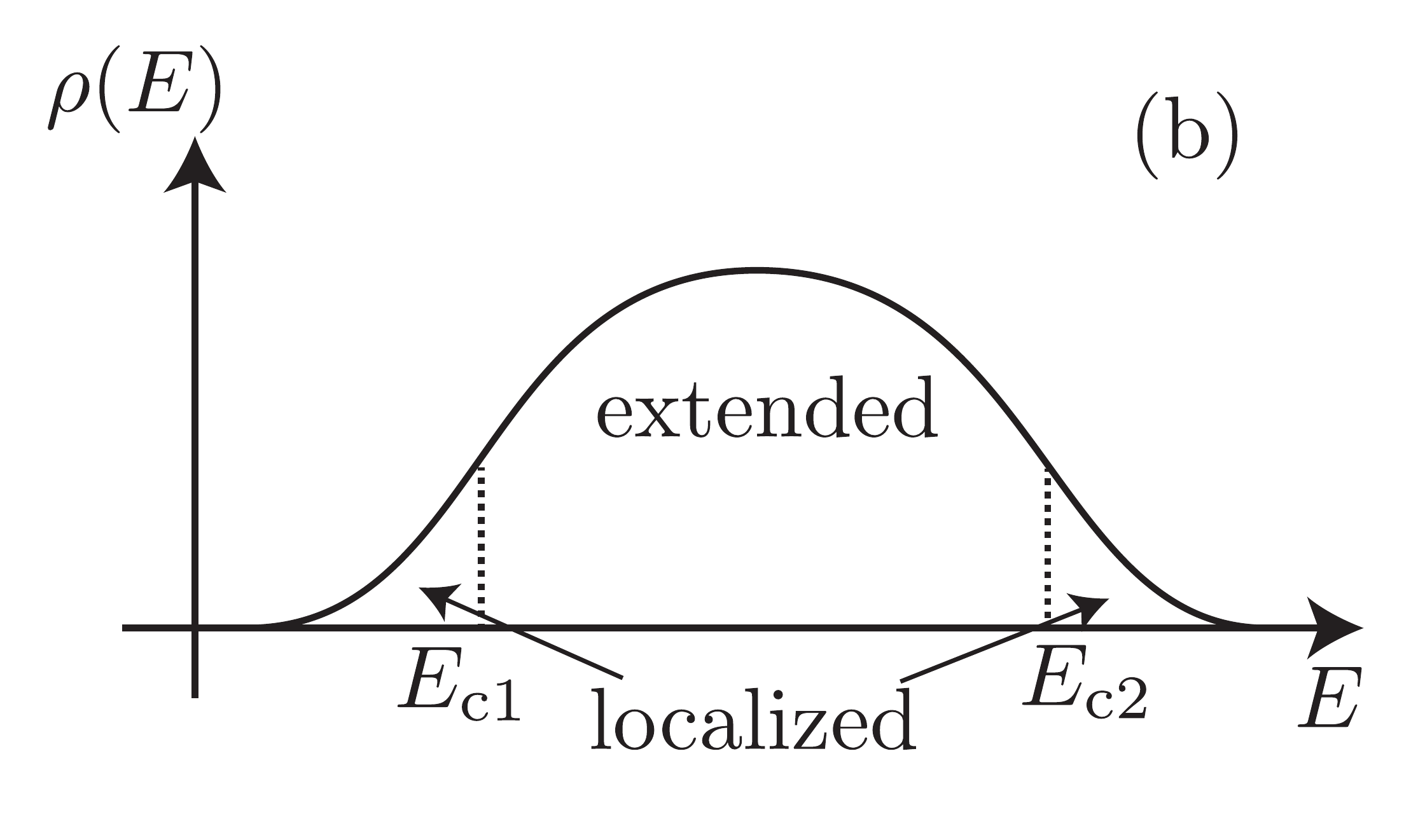}\\
    \includegraphics[width=6cm]{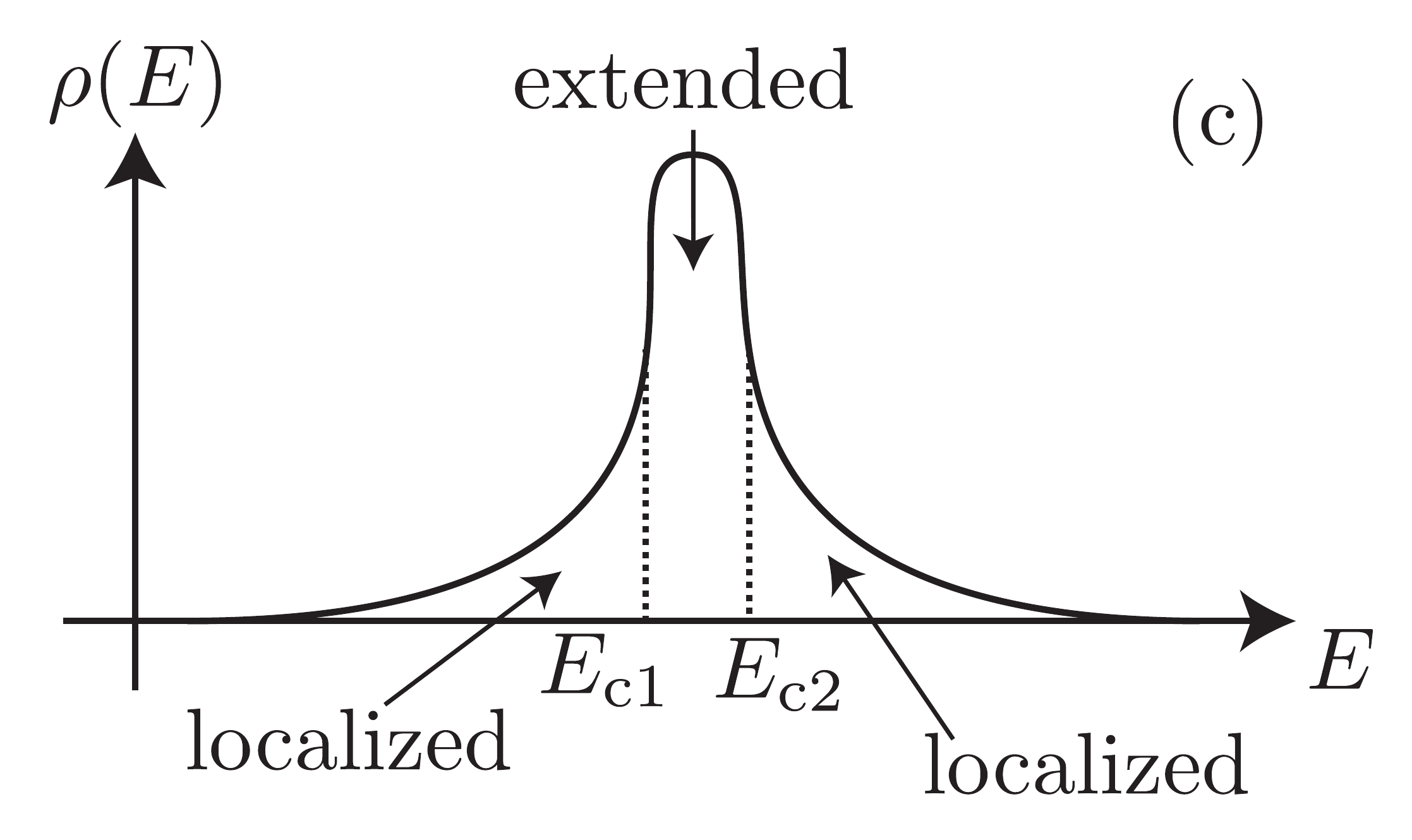}
    \caption{ Schematic pictures of the density of states, $\rho(E)$, as a function of the energy $E$ in three dimensions for (a) a nonrandom system, (b) with two mobility edges near the band edges, (c) with two mobility edges near the center of the band. }
    \label{DOS}
 \end{figure}

Clearly, if $\sigma^{}_0(E)$ is symmetric with respect to $\mu$  then $L_{12}=0$ and the Seebeck coefficient $S=0$. Therefore, the thermoelectric effect requires breaking the electron-hole symmetry.
An extreme case of such breaking arises when $\mu$ is close to a mobility threshold $E_c$, such that $\sigma^{}_0(E)=0$ for $E<E_c$ and $\sigma^{}_0(E)>0$ for $E>E_c$.
This led to many studies of thermoelectricity in systems which obey the leading power-law behavior,
\begin{align}
\sigma^{}_0 (E) = \begin{cases} 0 \ &(\text{for $E<E_c$}), \\
A|(E-E_c)/E_c|^x \ &(\text{for $E\geq E_c$}), \end{cases} \label{eq:0Tcond}
\end{align}
where $A$ is a system-dependent constant (with the dimensions of conductivity) and $x$ is a universal exponent (which depends on the dimension, $d$,  and on  the symmetry of the system). Such behavior  arises near  band edges of nonrandom systems, where the electronic density of states vanishes in the gap outside the band,  \cite{hicks} and near the  mobility edge of disordered electronic systems which undergo the  Anderson localization transition.\cite{lee1985,enderby1990,kramer1993,FSS}
In the former case, the density of states near the band edge in $d$ dimensions, with a quadratic dispersion $(E-E_c)\propto k^2$,  yields $x=(d-2)/2$ [see Fig.~\ref{DOS}(a)]. In the absence of electron-electron interactions, the Anderson localization transition exists only for $d>2$, and for $d=3$ numerical estimates yield $x=(d-2)\nu\approx 1.5$, where $\nu$ is the critical exponent for the localization length, \cite{ohtsuki1}
\begin{align}
\xi=\xi_0|(E-E_c)/E_c|^{-\nu}.
\end{align}

 Theoretical analyses  of Eqs.~(\ref{LL}) with Eq.~(\ref{eq:0Tcond})  are presented in quite a number of papers. \cite{cutler1969, sivan1986,kearney1988,castellani1988,kapitulnik1992,enderby1994,guttman1995,durczewski1998,villagonzalo1999,Imry2010,benenti2016}
Below we  comment on these analyses and add several new approximants for the thermoelectric coefficients. In addition, we include three new generalizations. First, we note that Eq.~(\ref{eq:0Tcond}) contains only the leading singularity in $\sigma^{}_0(E)$, very close to $E_c$.  Irrelevant variables near the localization fixed point introduce  corrections to this leading behavior for $E>E_c$, of the form \cite{FSS}
\begin{align}\label{corr}
\sigma^{}_0(E)=A|(E-E_c)/E_c|^x [1+a|(E-E_c)/E_c|^y+...],
\end{align}
where $a$ is the amplitude of the leading correction term and the dots represent higher-order corrections. For the three-dimensional Anderson localization the singular correction exponent $y$ seems to be much larger than $1$.\cite{FSS} Therefore, larger deviations  are expected to arise from   analytic corrections, with $y=1$, which may  result from nonlinear scaling fields \cite{AAlin} and from  the  energy dependence of the density of states. Section \ref{IS}  presents detailed calculations of  the thermoelectric coefficients in various regimes, including these corrections.

Second, the  finite-size dependence  of the Onsager coefficients  is introduced in Sec.~\ref{FSC}.   For a sample of linear size $L$, Eq.~(\ref{eq:0Tcond}) must be replaced by  the scaling form \cite{FSS}
\begin{align}
\sigma^{}_0(E,L)=A|(E-E_c)/E_c|^x{\cal F}(L/\xi),
\label{FSS}
\end{align}
where ${\cal F}(z)$ is a universal scaling function which obeys
\begin{align}
{\cal F}(u)=\begin{cases} 1 \ &(\text{for $u\gg 1,\ E>E_c$}), \\
u^{-x/\nu} \ &(\text{for $|u|\ll 1$}),\\
e^{-u} \ &(\text{for $u\gg 1,\ E<E_c$}). \end{cases} \label{FSS1}
\end{align}
In three dimensions  the exponents  are related by  $x/\nu=d-2=1$. \cite{gang4}

\begin{figure}
    \includegraphics[width=6cm]{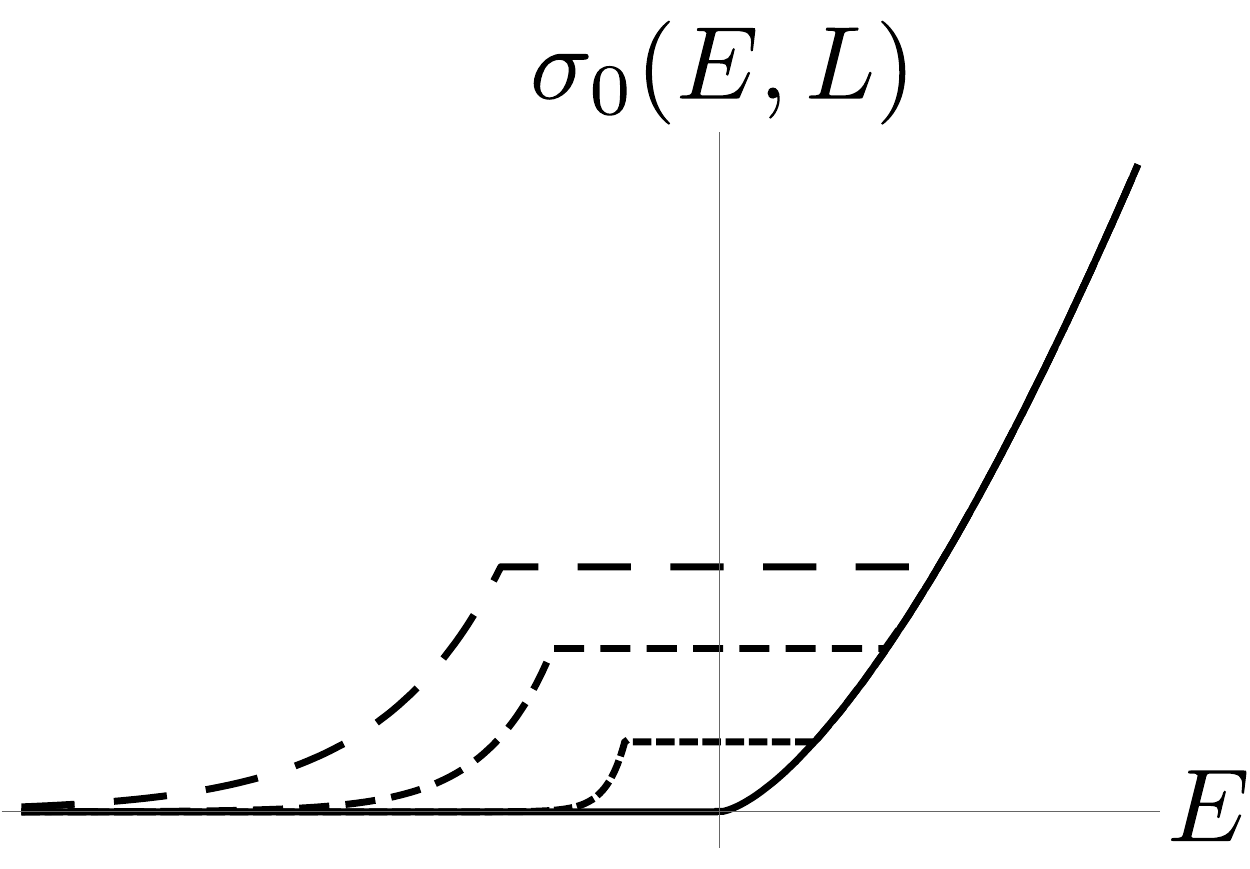}
    \caption{ A schematic picture  of the energy dependence of the zero-temperature conductivity  in a finite system. The flat horizontal line is broader and higher for smaller samples.}
    \label{Finitesigma}
 \end{figure}

 Figure \ref{Finitesigma}  illustrates  the zero-temperature conductivity of a finite system. The plateaus appear in the region $|(E-E_c)/E_c|<(L/\xi_0)^{-1/\nu}$, where $L<\xi$ and therefore the zero-temperature conductivity depends only on $L$ and not on $\xi$.  As the system size $L$ decreases, the width of the plateau increases. Since on this plateau $\sigma^{}_0$ does not depend on the energy, electron-hole symmetry is maintained and this region will not contribute to the Seebeck coefficient. Below we present new explicit results for the size dependence of the various thermoelectric coefficients and ratios.

Third, we note that  the above discussion assumed a single mobility edge at $E=E_c$, so that $\sigma_0(E)$ is nonzero for all energies  $E>E_c$, as in Eq.~(\ref{eq:0Tcond}). As noted in Refs.~\onlinecite{ziman1969}, \onlinecite{economou1970} and \onlinecite{economou1971},  in a band of a finite width, the  Anderson localization arises in the two band  tails, and therefore the zero-temperature conductivity is nonzero only over a finite energy range. For the nonrandom case [Fig.~\ref{DOS}(a)], the effects of the finite width of the energy bands was emphasized in Refs.~\onlinecite{dresselhaus} and \onlinecite{jeong}. However, we are not aware of any analysis of  thermoelectricity with two mobility edges  at $E_{c1}$ and $E_{c2}$ [Figs.~\ref{DOS}(b) and (c)].  For a large splitting between the two mobility edges, $E_{c2}-E_{c1} \gg \kB T$, it is enough to consider a single mobility edge [the function $F(E)$ in Eqs.~(\ref{LL}) ``captures'' only one mobility edge]. However, when the width $\kB T$ of $F(E)$ is larger than $E_{c2}-E_{c1}$,  both mobility edges  should be included.  If one assumes a symmetric band around $E=0$, with fully localized states for $|E|>|E_c|$ [Figs.~\ref{DOS}(b)  and \ref{DOS}(c)], then Eq.~(\ref{eq:0Tcond}) must be replaced by
\begin{align}
\sigma^{}_0 (E) = \begin{cases} 0 \ &(\text{for $|E|>|E_c|$}), \\
A[1-|E/E_c|]^x \ &(\text{for $|E|\leq |E_c|$}). \end{cases}
\label{band}
\end{align}
(The absolute value of $E$ was missing in Ref.~\onlinecite{villagonzalo1999}).\cite{com1} In Sec.~\ref{TM}  we extend the analysis to the case of such a narrow band.   When $|E_{c}|\gg \kB T$, the two localization thresholds are far apart, and the results of a single threshold   are reproduced.

However, when $|E_c|<\kB T$, the  zero-temperature conductivity is nonzero only over a narrow range, and then we find a large thermoelectric
efficiency. In the limit of a very narrow energy range, such an increase was originally noted by Mahan and Sofo.\cite{mahan,yamamoto2015} Since the width $2|E_c|$ decreases with increasing disorder [going from Fig.~\ref{DOS}(b) to \ref{DOS}(c)], we thus find that {\it increasing the disorder generates more efficient thermoelectricity.} \cite{power}

\section{The infinite system}
\label{IS}

\subsection{General considerations}

Defining
\begin{align}
\epsilon\equiv \frac{E-\mu}{\kB T}, \ \ \ \ z \equiv \frac{\mu-E_c}{\kB T},
\label{epz}
\end{align}
 and using Eq.~(\ref{corr}), we can rewrite Eqs.~(\ref{LL}) as
\begin{align}
L_{11} &= At^x \big[K_0(x,z)+at^yK_0(x+y,z)+\cdots\big],\nonumber\\
L_{12}&= A\frac{\kB T}{|e|}t^x \big[K_1(x,z)+at^yK_1(x+y,z)+\cdots\big],\nonumber\\
L_{22}&=  A\left(\frac{\kB T}{e}\right)^2 t^x \big[K_2(x,z)+at^yK_2(x+y,z)+\cdots\big],
\label{intL}
\end{align}
where
\begin{align}
t=\kB T/|E_c|.
\end{align}
The coefficient $A$ in Eq.~(\ref{intL}) comes from Eq.~(\ref{eq:0Tcond}) or from Eq.~(\ref{corr}); it is canceled in the expressions for $S$, $ZT$, and ${\cal L}$, which we analyze below. The coefficient  $a$ comes from the leading correction in Eq.~(\ref{corr}), and  the dots denote terms of order $(at^y)^2$ and higher. The functions $K_n(x,z)$ are defined as
\begin{align}\label{KK}
K_n(x,z)=\int_{-z}^\infty d\epsilon\epsilon^n(\epsilon+z)^x\frac{1}{4\cosh^2(\epsilon/2)}.
\end{align}

\begin{figure}
  \begin{center}
    \includegraphics[width=7cm]{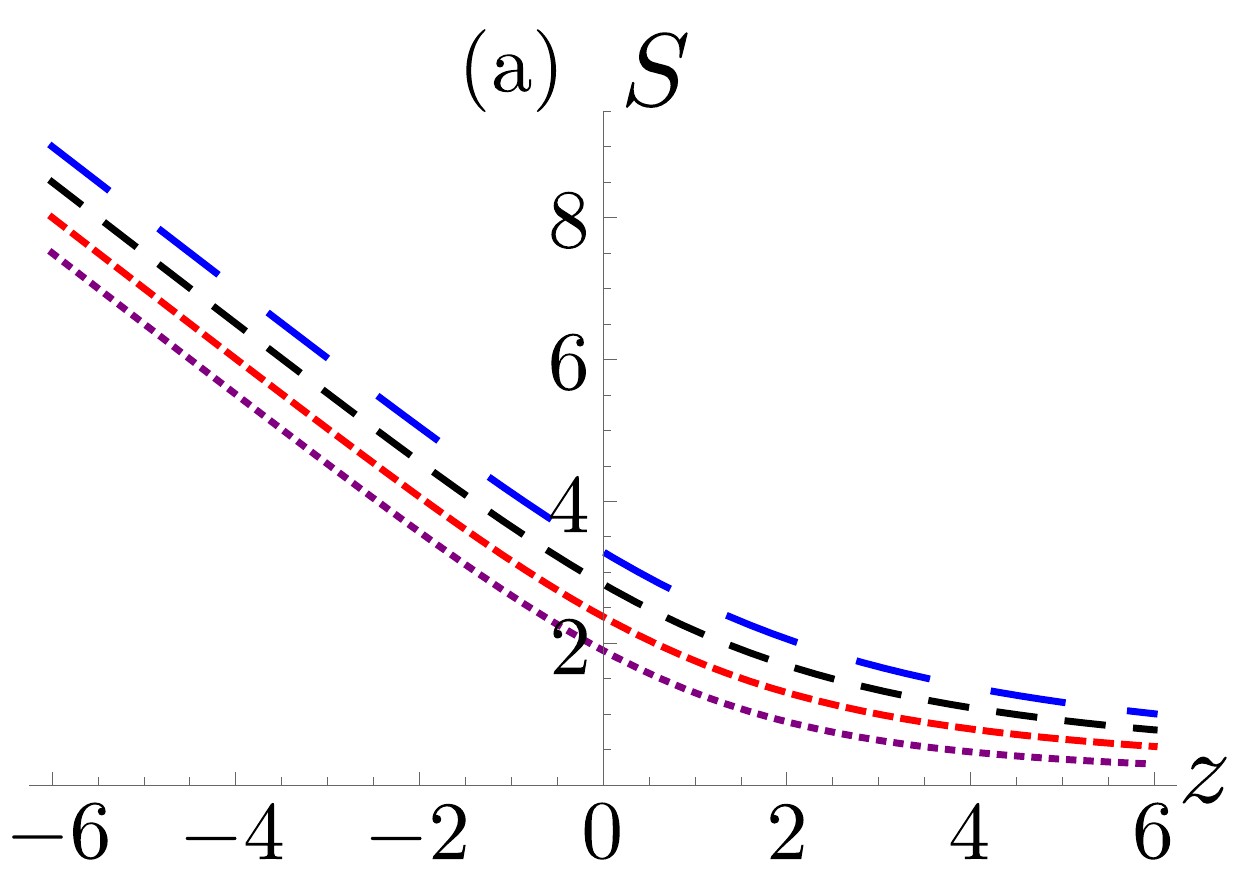}\\
    \vspace{\baselineskip}
    \includegraphics[width=7cm]{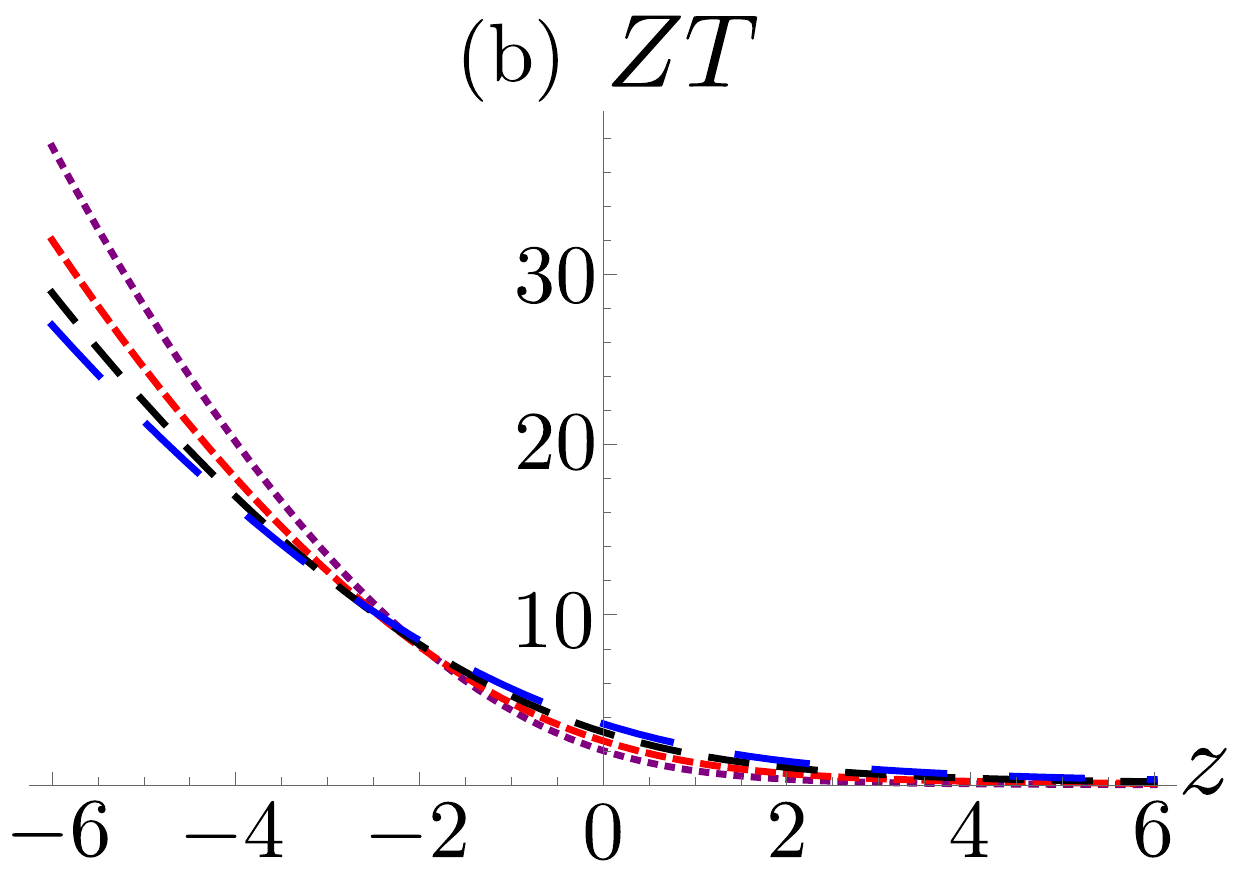}\\
    \vspace{\baselineskip}
    \includegraphics[width=7cm]{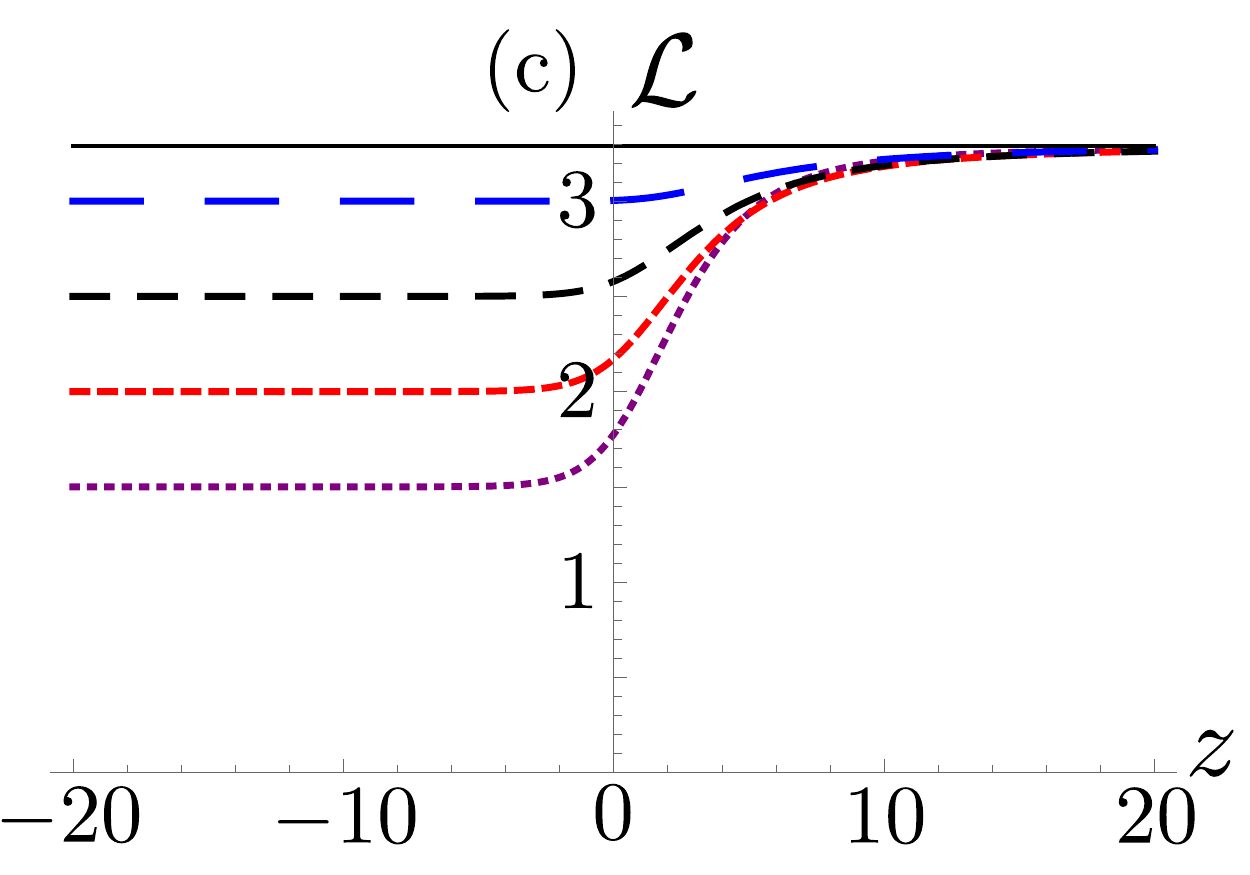}
     \caption{The leading universal dependences of the Seebeck coefficient $S$ [in units of $\kB/|e|$, panel (a)], the figure of merit $ZT$ [panel (b)], and the Wiedemann-Franz ratio ${\cal L}$ [in units of $(\kB/|e|)^2$, panel (c)] versus $z=(\mu-E_c)/(\kB T)$, Eqs.~(\ref{plotS}), (\ref{plotZT}) and (\ref{plotL}), with $x=0.5$ [dotted (magenta) curve], $x=1$ [small-dashed (red) curve], $x=1.5$ [medium-dashed (black) curve], and $x=2$ [large-dashed (blue) curve]. }
    \label{fig1}
  \end{center}
 \end{figure}

 In terms of the integrals (\ref{KK}),  the Seebeck coefficient $S$, the figure of merit $ZT$, and the Wiedemann-Franz ratio, ${\cal L}$, take the forms
\begin{widetext}
\begin{align}
S=\frac{\kB}{|e|}\frac{K_1(x,z)}{K_0(x,z)}\left[1+at^y\left(\frac{K_1(x+y,z)}{K_1(x,z)}-\frac{K_0(x+y,z)}{K_0(x,z)}\right)+\cdots\right],
\label{SSSS}
\end{align}
\begin{align}
ZT&=\frac{K_1(x,z)^2}{K_0(x,z)K_2(x,z)-K_1(x,z)^2}\left[1+at^y\left(2\frac{K_1(x+y,z)}{K_1(x,z)} \right. \right.\nonumber\\
&\left.\left. -\frac{K_0(x,z)K_2(x+y,z)+K_2(x,z)K_0(x+y,z)-2K_1(x,z)K_1(x+y,z)}{K_0(x,z)K_2(x,z)-K_1(x,z)^2}\right)+\cdots\right]
\label{ZTZT}
\end{align}
and
\begin{align}
{\cal L}&=\left(\frac{\kB}{e}\right)^2\frac{K_0(x,z)K_2(x,z)-K_1(x,z)^2}{K_0(x,z)^2}\biggl[1+\nonumber\\
&at^y\left. \left(\frac{K_0(x,z)K_2(x+y,z)+K_2(x,z)K_0(x+y,z)-2K_1(x,z)K_1(x+y,z)}{K_0(x,z)K_2(x,z)
-K_1(x,z)^2}-2\frac{K_0(x+y,z)}{K_0(x,z)}\right)+\cdots\right].
\label{LLLL}
\end{align}
\end{widetext}
 Using only the leading term in the zero-temperature conductivity [Eq.~(\ref{eq:0Tcond})],  Eqs.~(\ref{SSSS}), (\ref{ZTZT}), and (\ref{LLLL}) become
\begin{align}
S=\frac{\kB}{|e|}\frac{K_1(x,z)}{K_0(x,z)},
\label{plotS}
\end{align}
\begin{align}
ZT=\frac{K_1(x,z)^2}{K_0(x,z)K_2(x,z)-K_1(x,z)^2},
\label{plotZT}
\end{align}
and
\begin{align}
{\cal L}&=\left(\frac{\kB}{e}\right)^2\frac{K_0(x,z)K_2(x,z)-K_1(x,z)^2}{K_0(x,z)^2},
\label{plotL}
\end{align}
 respectively. For a given value of the exponent $x$, these are  {\bf universal functions} of  $z=(\mu-E_{c})/(k_{\rm B}T)$ [Eq.~(\ref{epz})]. These functions are  calculated numerically and plotted in Fig.~\ref{fig1}  for four values of $x$.  Similar numerical plots for $S$ appeared in Refs.~\onlinecite{villagonzalo1999}  and \onlinecite{enderby1994} for  any value of $z$, and in Refs.~\onlinecite{Imry2010} and \onlinecite{benenti2016} for $z\geq 0$. Ones for $ZT$ appeared in Refs.~\onlinecite{durczewski1998} and \onlinecite{benenti2016} for  $z>0$.

As  seen in Fig.~\ref{fig1}(a), $S$ vanishes at positive infinite $z$ and increases monotonically to infinity as $z\rightarrow -\infty$.
Also, at each value of $z$ the Seebeck coefficient increases with $x$.  Therefore in three dimensions we expect this coefficient to be larger near the Anderson threshold in the random system ($x=\nu\approx 1.5$), compared to its behavior near the band edge  of the normal metal, for which $x=0.5$.

 The figure of merit  $ZT$ [Fig.~\ref{fig1}(b)] also vanishes at infinite $z$ and increases monotonically to infinity as $z\rightarrow -\infty$. Although $ZT$ increases monotonically with $x$ for $z>0$, the lines cross at negative $z$ and $ZT$ decreases with increasing $x$ for large negative $z$.
Interestingly, $ZT$ crosses the ``desired" value of 3 already at $z\approx -0.5$.

 As seen in Fig.~\ref{fig1}(c),  ${\cal L}$ always approaches the Lorenz value [${\cal L}_0=(\pi^{2}/3)\kB^2/e^2$, shown in the figure by the thin horizontal line] at large $z$, but it decreases with decreasing $x$, approaching different $x$-dependent constants for large negative values of  $z$.

\subsection{Approximations}
\label{approx}

Appendix \ref{AppA}  elaborates on the computation of   the integrals $K_n(x,z)$,  Eq.~(\ref{KK}), in three regimes of $z$, i.e. $z\gg 1$, $|z|\ll 1$, and $z\ll -1$, and uses them to obtain analytic approximations for the  Onsager linear-response coefficients $L_{ij}$. The behavior of the Seebeck coefficient $S$,  the figure of merit $ZT$, and the Wiedemann-Franz ratio ${\cal L}$ in these regimes is discussed in the following. As can be seen in Fig.~\ref{app}, the three approximants found in the Appendix are accurate over wide ranges of $z$.

\begin{figure}
  \begin{center}
    \includegraphics[width=7cm]{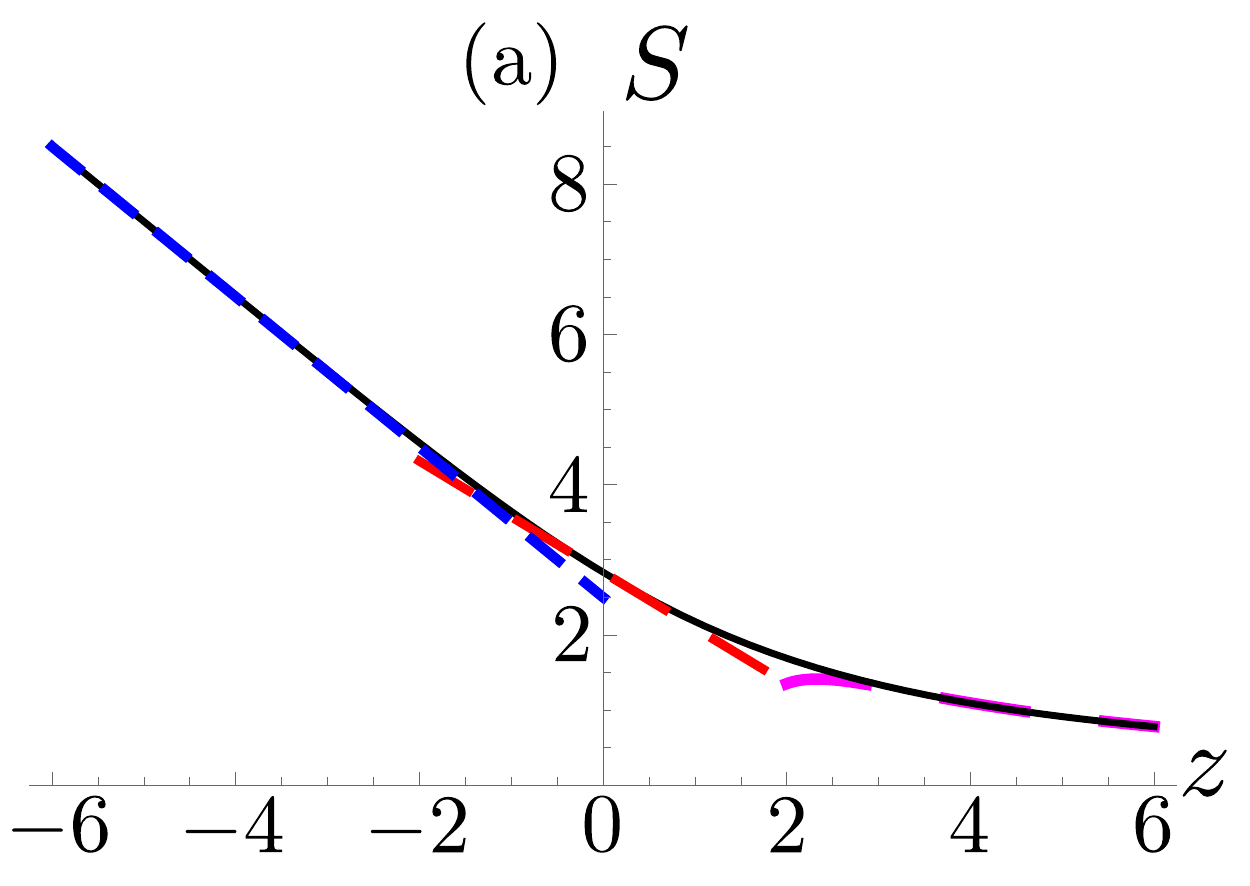}\\
    \vspace{\baselineskip}
    \includegraphics[width=7cm]{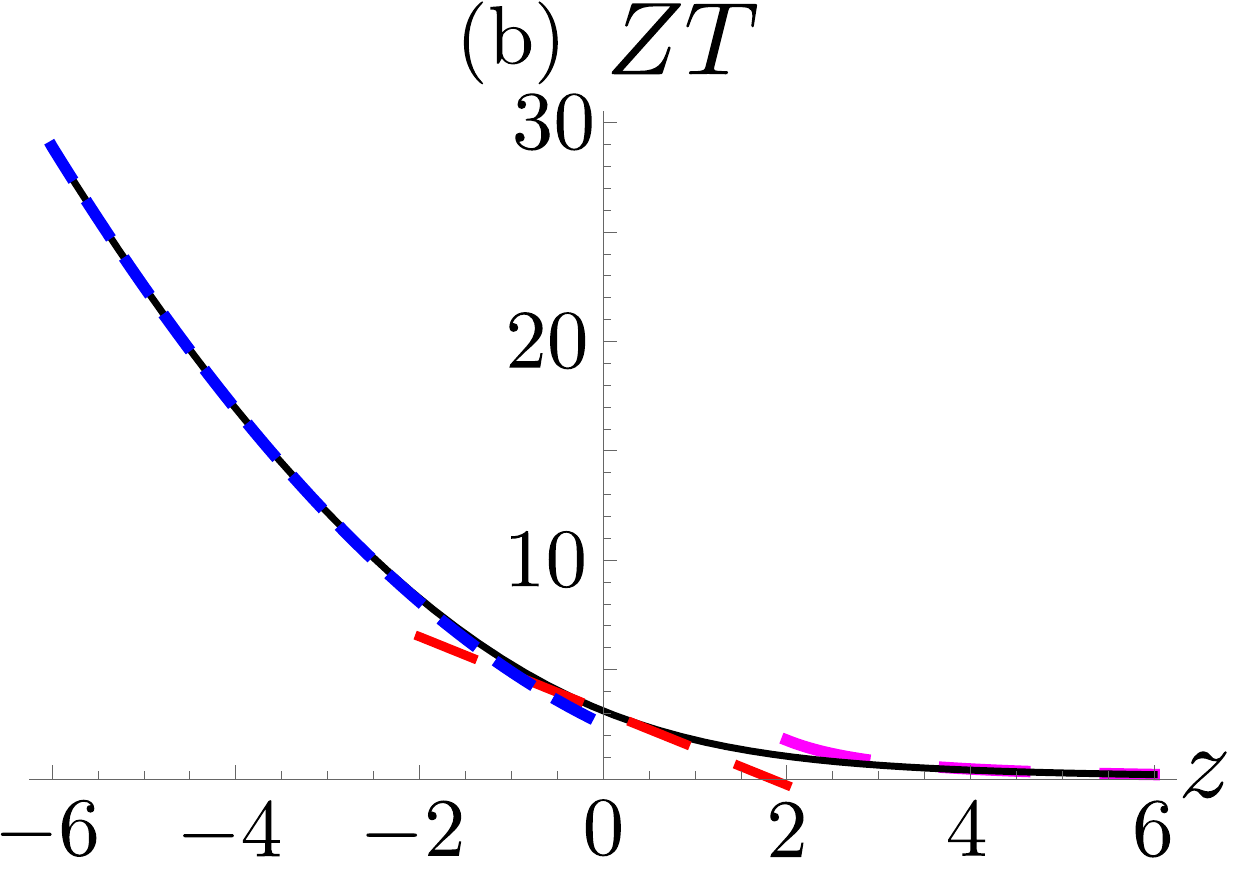}\\
    \vspace{\baselineskip}
     \includegraphics[width=7cm]{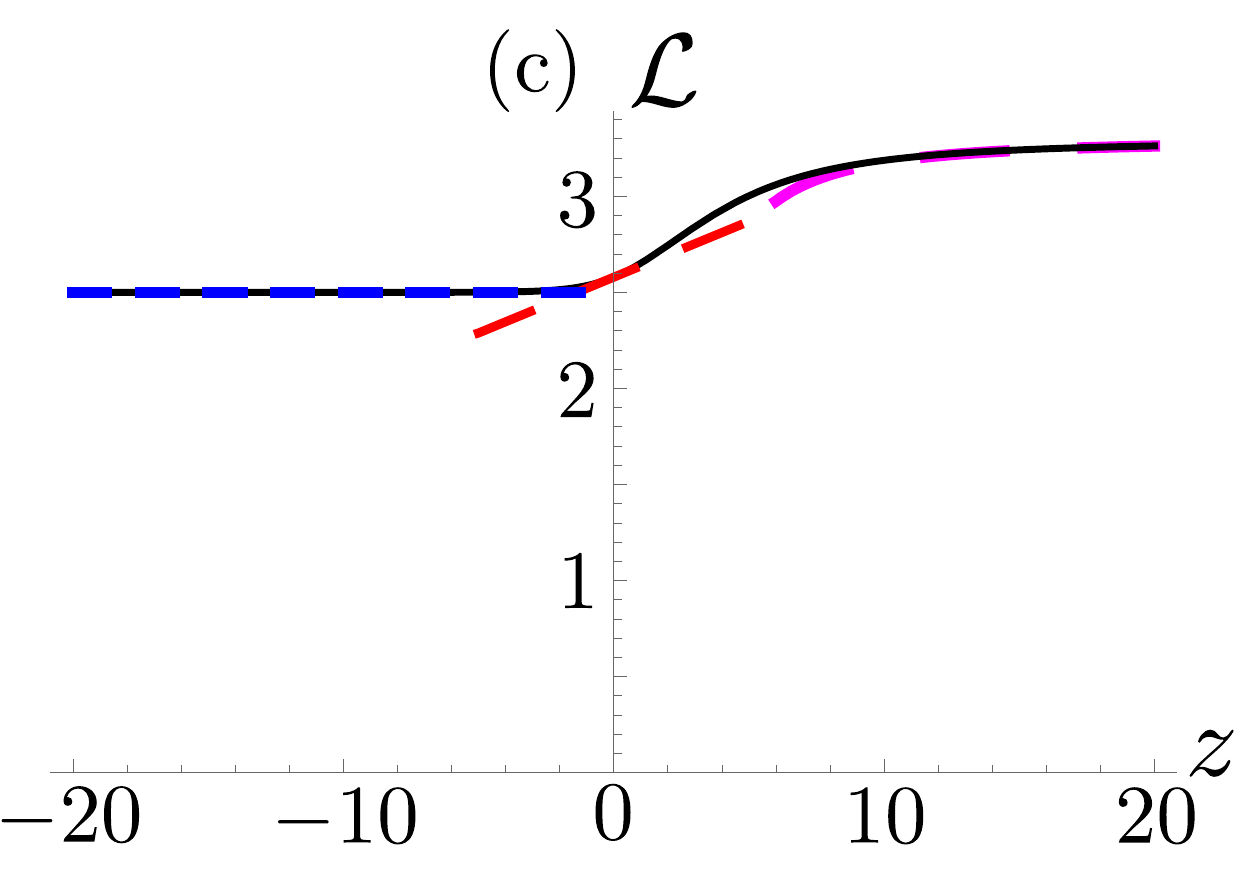}
    \caption{Comparisons of the full curves from Fig.~\ref{fig1}, with $x=1.5$ [solid (black) curves], with the various approximants (dashed lines). Each panel shows two leading terms in Eqs.~(\ref{SlowT}), (\ref{ZTlowT}), or (\ref{LlowT}) for $z\gg 1$ [larger-dashed (magenta) curve] and in Eqs.~\eqref{ShighT1}, \eqref{ZThighT1} or (\ref{WLhighT}) for $|z|\ll 1$ [dashed (red) line], and one leading term in Eqs.~(\ref{Snegz}), (\ref{ZTnegz}), or (\ref{LLnegz})  for $z\ll -1$ [dotted (blue) line].    }
    \label{app}
  \end{center}
 \end{figure}

\vspace{.5cm}
\noindent\underline{1. $z\gg 1$}: For $\mu-E_c\gg \kB T$, i.e., $z\gg 1$, the expansion (\ref{A2}) in powers of $1/z$ is equivalent to the Sommerfeld approximation. \cite{ashcroft}
The leading term in $S$ decays for $(\mu-E_c)\gg \kB T$ as \cite{sivan1986,kearney1988,enderby1994,villagonzalo1999,Imry2010,benenti2016}
\begin{align}
S\approx\frac{\kB}{|e|}\frac{\pi^2}{3}\frac{x}{z}=\frac{\kB}{|e|}\frac{\pi^2}{3}\frac{x\kB T}{\mu-E_c}.
\label{eq:SlowT}
\end{align}
 An accurate measurement of the decay of $S$ for $\mu-E_c\gg \kB T$ can therefore yield the value of the exponent $x$. \cite{Imry2010}

Equation (\ref{SlowT}) contains two leading corrections to the leading-order Seebeck coefficient, Eq.~(\ref{eq:SlowT}).
The first correction, $\pi^2x(x-1)(x-7)/(15z^2)$, is universal. It  arises  from the expansion of $S$ to the second order in $1/z$ \cite{Imry2010}   and  modifies the linear  temperature dependence of $S$ by adding a term of order $T^3$. Writing
\begin{align}
\frac{3|e|}{\kB \pi^2}S=c^{}_1T+c^{}_3T^3, \label{eq:SlowT2}
\end{align}
one   finds
\begin{align}
\frac{c^{}_3}{c_1^2}=\frac{\pi^2}{15}x(x-1)(x-7).
\end{align}
 Interestingly, this ratio becomes negative for $x>1$. Measuring this ratio can yield another identification of the exponent $x$.
As seen in Fig.~\ref{app}(a), the approximation Eq.~(\ref{SlowT}) (at $a=0$)  is excellent for $z>3$.

The second correction to $S$ in Eq.~(\ref{SlowT}), $ayt^y$, that comes from the leading correction to scaling in Eq.~(\ref{corr}), introduces a nonuniversal  temperature dependence to the universal amplitude $(\pi^2/3)x \kB/|e|$ in Eq.~(\ref{eq:SlowT}). At a fixed chemical potential $\mu$, this correction also implies that $S$ is not linear in the temperature: $S\propto T[x+ay(\kB T/|E_c|)^y]$. (Note that  the  temperature dependence of $\mu$, which is quadratic in $T$,  is ignored. \cite{ashcroft,villagonzalo1999}) 
Plotting $S/T$ versus $T$ could help identify the leading correction exponent $y$. More on this correction  below.

The leading term in $ZT$ decays for $\mu-E_c\gg \kB T$ as \cite{kapitulnik1992,benenti2016}
\begin{align}
ZT\approx\frac{\pi^2}{3}\frac{x^2}{z^2}=\frac{\pi^2}{3}\left(\frac{x\kB T}{\mu-E_c}\right)^2.
\label{eq:ZTlowT}
\end{align}
Equation (\ref{ZTlowT})  contains the  two leading corrections to Eq.~(\ref{eq:ZTlowT}). The first (universal) correction modifies the quadratic temperature dependence of $ZT$, adding a new term of order $T^4$. As seen on the right-hand side of Fig.~\ref{app}(b), including this new term gives an excellent approximation for $z>3$.
The second correction, whose origin is the leading correction in Eq.~(\ref{corr}), introduces a nonuniversal correction to the universal amplitude $\pi^2x^2/3$ in Eq.~(\ref{ZTlowT}), which varies with $t=\kB T/|E_c|$. 

For $z\gg 1$, Eq.~(\ref{LlowT}) shows the decrease of ${\cal L}$ from the universal Lorenz value ${\cal L}_0$.  Interestingly enough, this $x$-dependent expression has not appeared in the literature (even for nonrandom band edges). Unlike the $x$-independent ${\cal L}_0$, this correction term, of order $1/z^2\propto T^2$, does depend on $x$. Note that the correction of order $at^y$ vanishes.

\vspace{.5cm}

\begin{figure}
   \includegraphics[width=3.6cm]{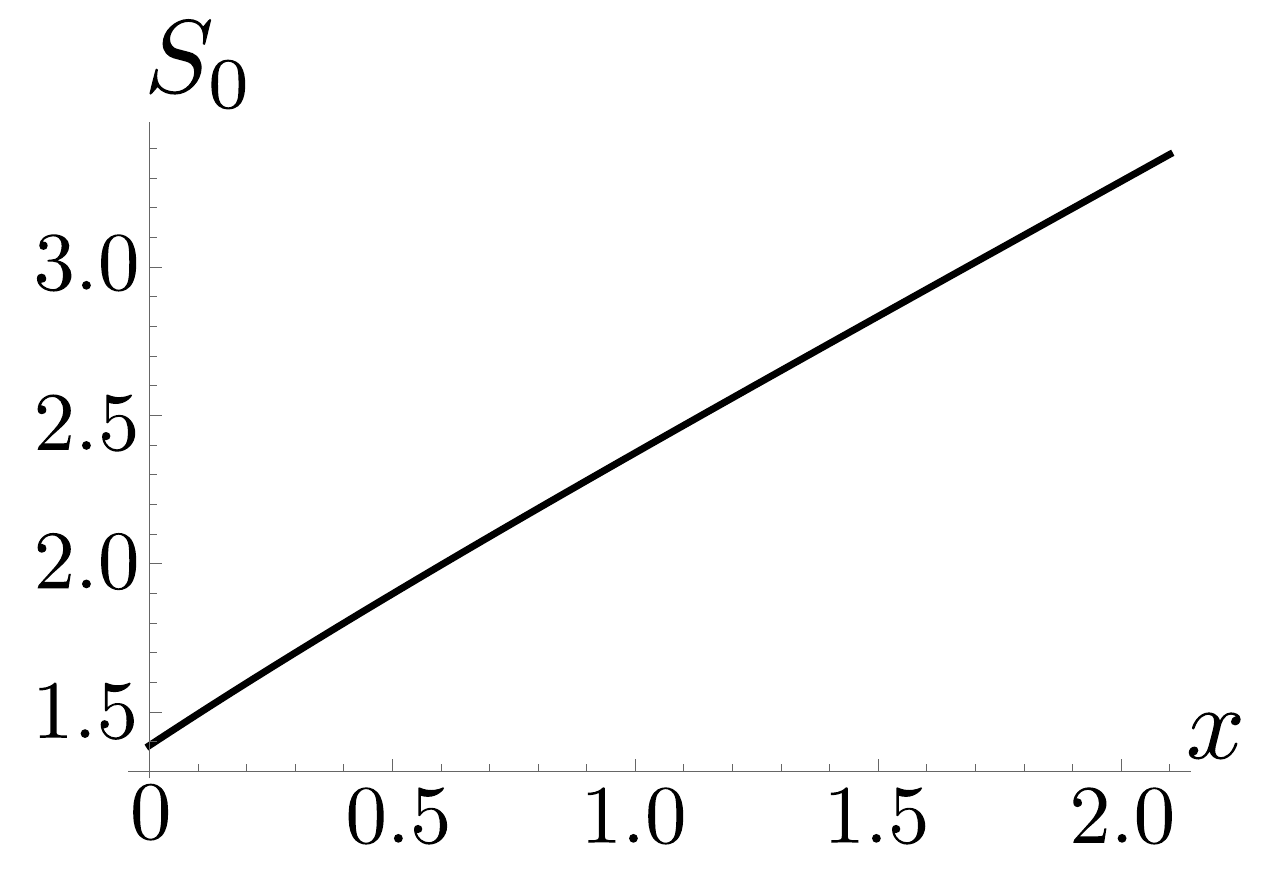}\ \ \
   \includegraphics[width=3.6cm]{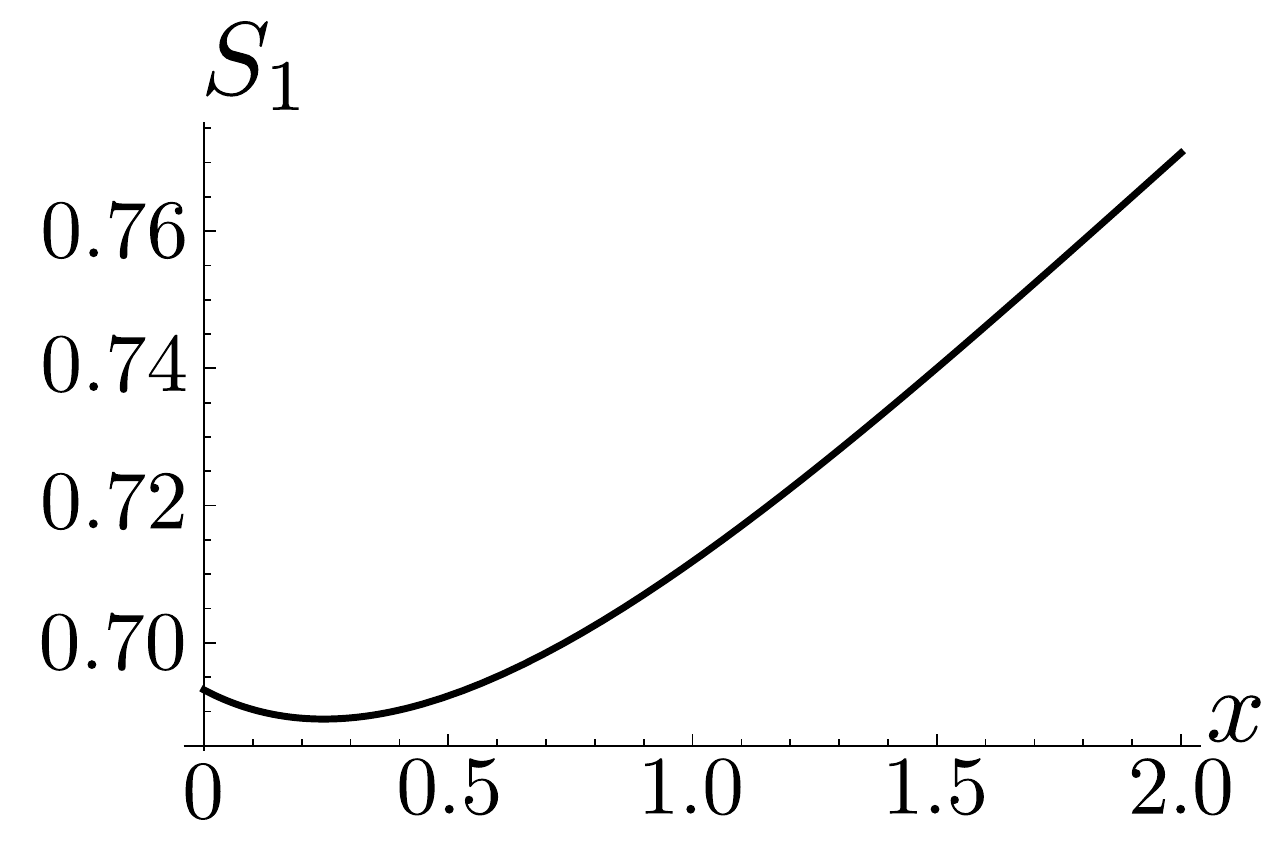}
    \caption{The $x$ dependences of the coefficients $S_0$ (left) and $S_1$ (right) (in units of $\kB/|e|$) in Eq.~(\ref{ShighT}).}
    \label{Svsx}
 \end{figure}

\noindent\underline{2. $|z|\ll 1$}:  If one ignored the constraint $z\gg 1$, i.e., $\kB T\ll \mu-E_c$ in Eq.~(\ref{eq:SlowT}), then one would wrongly conclude  an apparent divergence of  the Seebeck coefficient $S$ for $\mu\rightarrow E_c^+$ at fixed $T$.
This led to some confusion in early papers based on the Sommerfeld approximation  (e.g.,  Ref.~\onlinecite{castellani1988}). However, the situation was clarified in Refs.~\onlinecite{villagonzalo1999} and \onlinecite{enderby1994}, which showed that {\it at} the mobility edge $\mu=E_c$ (i.e., $z=0$), $S$ approaches a {\it finite universal} value, which follows from the leading terms in Eq.~(\ref{LhighT}):
\begin{align}
S(z=0)=\frac{\kB}{|e|}\frac{(x+1)I_{x+1}}{xI_x},
\label{Sz0}
\end{align}
where $I_u$ is given in Eq.~(\ref{IInu}).  The  left panel in  Fig.~\ref{Svsx} shows $S_0=S(z=0)$ versus $x$. The thermopower increases monotonically with $x$ at $z=0$. Measuring $S$ at the mobility edge, $\mu=E_c$, can thus identify the exponent $x$.

 The first (universal) correction gives a linear dependence on $z=(\mu-E_c)/(\kB T)$, for $|z|\ll 1$; see Eq.~(\ref{ShighT}). Similar approximations, of the form $S=S_0-S_1z$, but with wrong coefficients, appeared in Refs.~\onlinecite{villagonzalo1999}, \onlinecite{sivan1986}, and \onlinecite{Imry2010}. Reference \onlinecite{kearney1988} gave a linear expression, without specifying the coefficients $S_0$ and $S_1$.  The $x$ dependences of the coefficients $S_0$ and $S_1$,  calculated from  Eq.~(\ref{ShighT}),  are displayed in Fig.~\ref{Svsx}.   As seen in Fig.~\ref{app}(a), the linear dependence of $S$ on $z$ fits the full curve reasonably well for $|z|<1/2$. Measuring the slope of this curve for small $z$ also yields information on the exponent $x$.
The second correction comes from the leading correction to scaling. This correction introduces a small nonuniversal nonlinear variation of $S$ with $t=\kB T/|E_c|$.

\begin{figure}[h]
    \includegraphics[width=4cm]{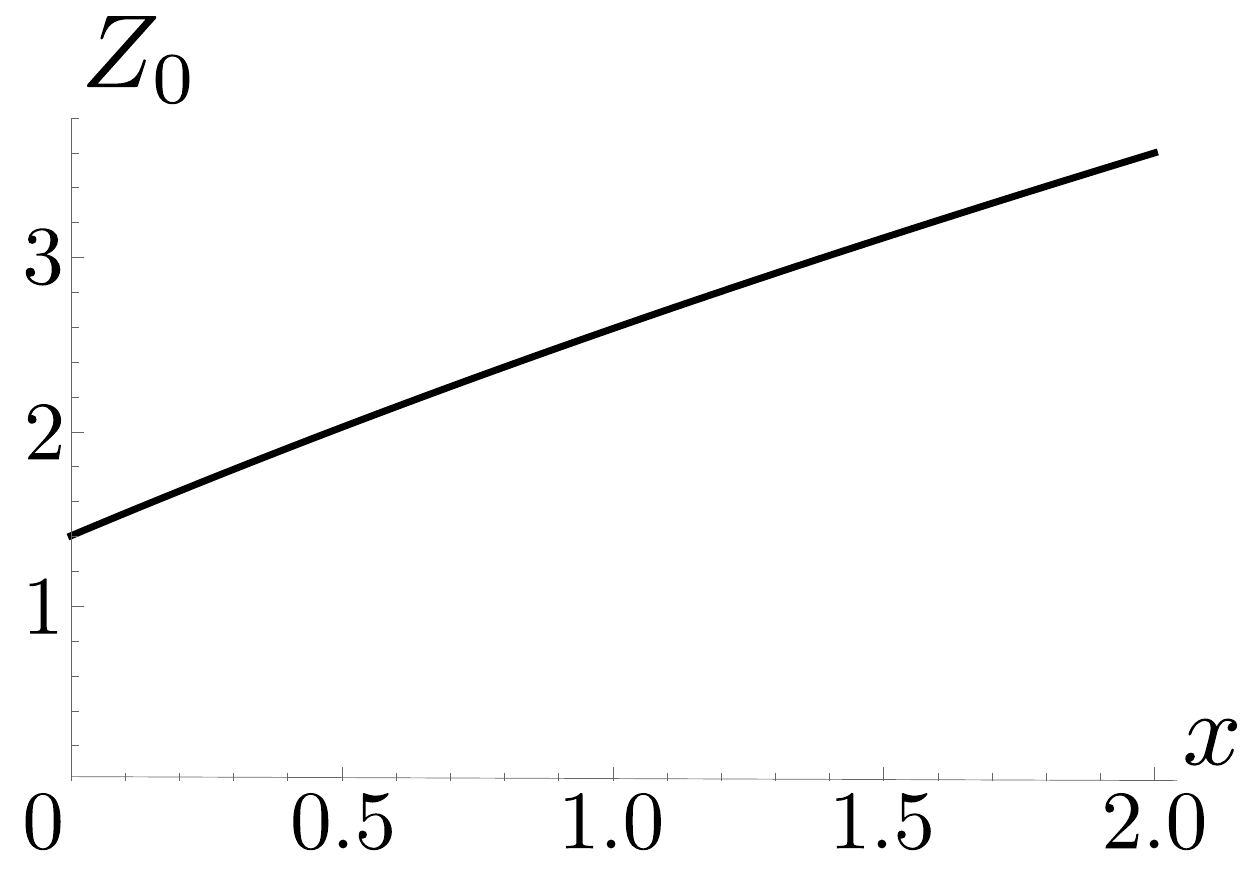}\ \ \
    \includegraphics[width=4cm]{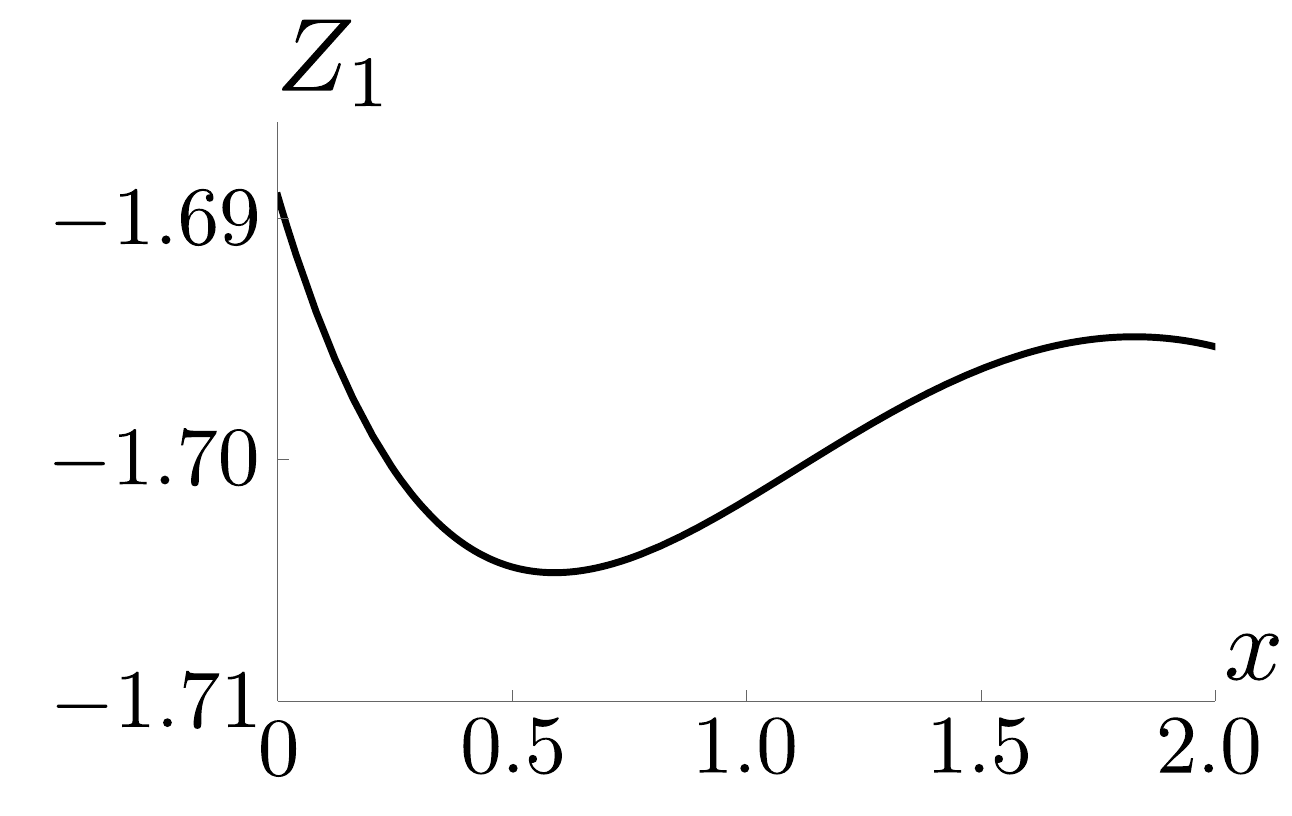}
    \caption{The  $x$-dependences of the coefficients $Z_0$ and $Z_1$ in $ZT\approx Z_0+Z_1z$ for $|z|\ll 1$,  Eq.~(\ref{ZThighT}).}
    \label{ZTvsx}
\end{figure}

Similarly, ignoring the constraint $\kB T\ll \mu-E_c$ in  Eq.~(\ref{eq:ZTlowT}) one would also find an apparent divergence of $ZT$ for $\mu\rightarrow E_c$ at fixed $T$.  
However,  the leading terms in Eq.~(\ref{LhighT}) imply that {\it at} the mobility edge $\mu=E_c$, $ZT$ approaches a finite universal value,
\begin{align}
&ZT(z=0) =Z_0
= \frac{(x+1)^2I_{x+1}^2}{(x+2)xI_{x+2}I_{x}-(x+1)^2I_{x+1}^2}. \label{eq:ZTz0}
\end{align}
The left panel of Fig.~\ref{ZTvsx} shows $Z_0=ZT(z=0)$ versus $x$, Eq.~(\ref{eq:ZTz0}).  A few specific values were listed in Ref.~\onlinecite{durczewski1998}.
As already seen in Fig.~\ref{fig1}(b), the figure of merit increases monotonically with $x$.

For $|z|\ll 1$, Eqs.~(\ref{ZThighT}) also give the leading corrections to the figure of merit.
The first (universal) correction practically gives a linear dependence on $z=(\mu-E_c)/(\kB T)$, for $|z|\ll 1$.  Equation (\ref{ZThighT}) is used to plot the  $x$-dependence of $Z_0$ and of $Z_1$, see  Fig.~\ref{ZTvsx}. Note that $Z_1$ is practically independent of $x$. The figure of merit decreases monotonically with increasing $z$. As seen in Fig.~\ref{app}(b), the linear dependence of $ZT$ on $z$ fits the full curve (for $a=0$) reasonably well for $|z|<1/2$.
The second correction, which comes from the leading correction to Eq.~(\ref{corr}), introduces a small temperature-dependent nonuniversal nonlinear variation of $ZT$ with $t=\kB T/|E_c|$.

\vspace{.5cm}

\noindent\underline{3. $z\ll -1$}: Finally, we turn to the ``insulating" side of the mobility edge, $\mu<E_c$. Although   the Onsager coefficients (including the conductivity) vanish in this regime at $T=0$, they become nonzero for $T>0$ due to the tail of the Fermi function. 
This regime, which apparently has not  attracted much previous interest, deserves a close inspection.
Equations (\ref{Lnegz}) show that   the Onsager coefficients decay exponentially in $z$ for $z\ll -1$. However, as seen from Figs.~\ref{fig1}(a) and \ref{fig1}(b), the Seebeck coefficient and the figure of merit grow significantly as $z$ becomes more negative.  Although Refs.~\onlinecite{kapitulnik1992} and \onlinecite{guttman1995} mention that $ZT$ may be large for negative $z$,    no explicit expression for this increase is provided.

Equations (\ref{Snegz}) and (\ref{ZTnegz}) give analytic approximations for $S$ and $ZT$ for $z\ll -1$.
Figure \ref{app}(a) shows the lowest-order (universal) term in Eq.~(\ref{Snegz}), which is linear in $z$. Clearly, this linear term represents an excellent approximation for $S$ for $z<-3$. The intercept of this straight line with the $S$ axis gives $(1+x)$.
Similarly, Fig.~\ref{app}(b) shows the lowest-order (universal) term of the figure of merit, $ZT=(1+x-z)^2/(1+x)$, which is quadratic in $z$. This quadratic term represents an excellent approximation of $ZT$ for $z<-2$. The intercept of this parabola with the $ZT$-axis, equal to  $(1+x)$, can be used to determine the value of the exponent $x$.
For $z\ll -1$, Eqs.~(\ref{Lnegz}) also yield
\begin{align}
{\cal L}=(\kB/e)^2\big[x+1+{\cal O}(e^z,at^y)\big],
\label{LLnegz}
\end{align}
which approaches the constant limit $(\kB/e)^2(x+1)$ [see Fig.~\ref{fig1}(c)].

\begin{figure}
    \includegraphics[width=5cm]{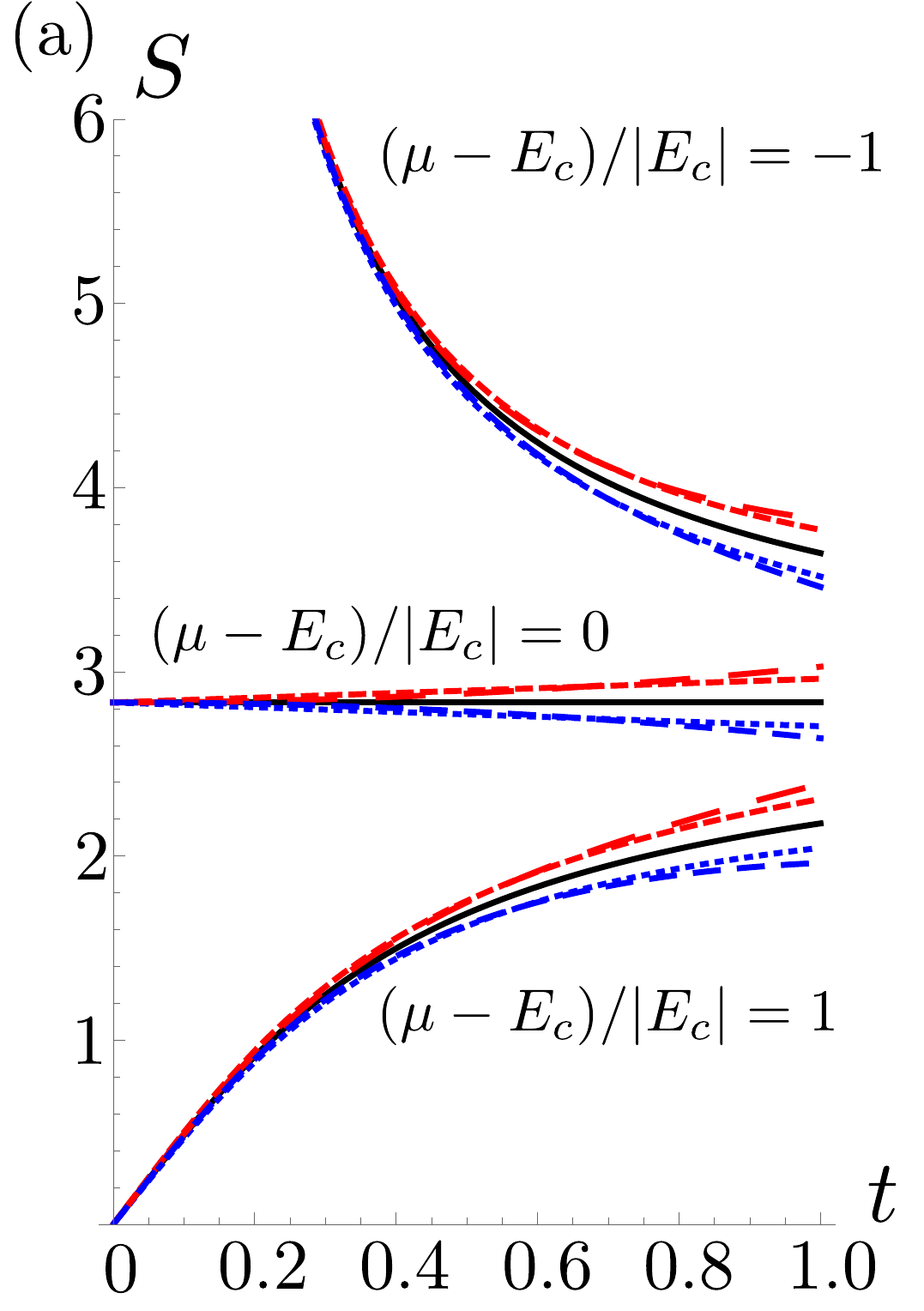}\\
    \vspace{\baselineskip}
    \includegraphics[width=5cm]{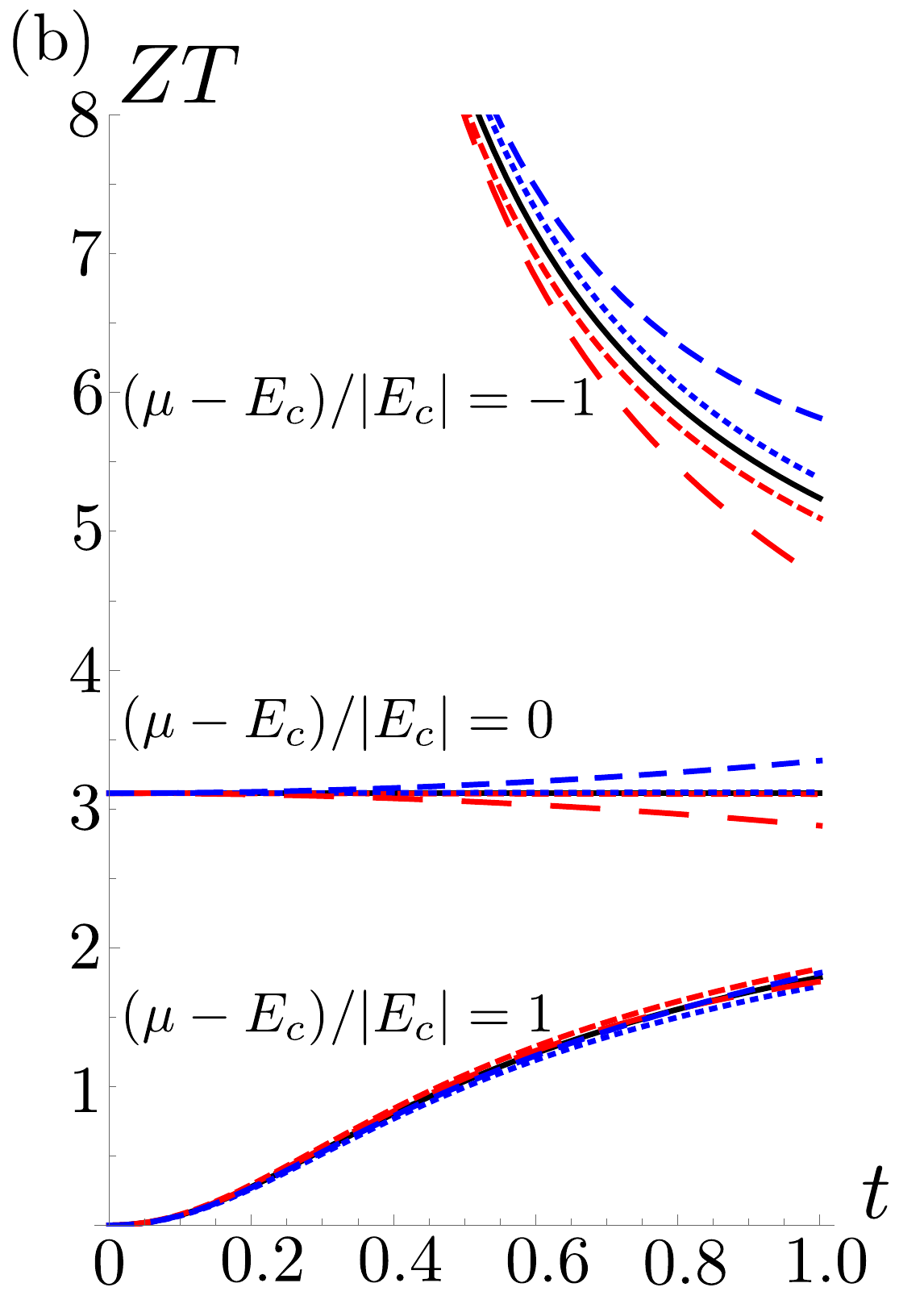}
    \caption{(a) The Seebeck coefficient in Eq.~\eqref{SSSS} and (b) the figure of merit, Eq.~(\ref{ZTZT}), versus $t = \kB T/|E_c|$ for $x=1.5$, and $y=1$, with $a=0$ [solid (black) curves], $a=0.05$ [dashed (red) curves], and $a=-0.05$ [dotted (blue) curves], as well as for $y=2$, with $a=0.01$ [largest-dashed (red) curves] and $a=-0.01$ [medium-dashed (blue) curves]: $(\mu-E_c)/|E_c|=-1$ (top), $(\mu-E_c)/|E_c|=0$ (middle), $(\mu-E_c)/|E_c|=1$ (bottom).
     }
    \label{S0}
 \end{figure}

As already stated, the nonuniversal correction of order $at^y$ introduces an additional small temperature-dependent nonlinear variation of $S$ with $t=\kB T/|E_c|$. Figure \ref{S0}  shows  the Seebeck coefficient, Eq.~(\ref{SSSS}), and the figure of merit, Eq.~(\ref{ZTZT}), for several values of $\mu-E_c$, $y$, and $a$.  Increasing  the temperature  increases the Seebeck coefficient for $a>0$ and decreases it for $a<0$. The effect is much stronger for larger $y$.
For $\mu-E_c >0$, increasing temperature  increases $ZT$ for $a>0$ and decreases it for $a<0$. For $\mu-E_c \leq 0$, increasing temperature  decreases $ZT$ for $a>0$ and increases it for $a<0$. Again, the effect is much larger for larger $y$.

\begin{figure}
    \includegraphics[width=5cm]{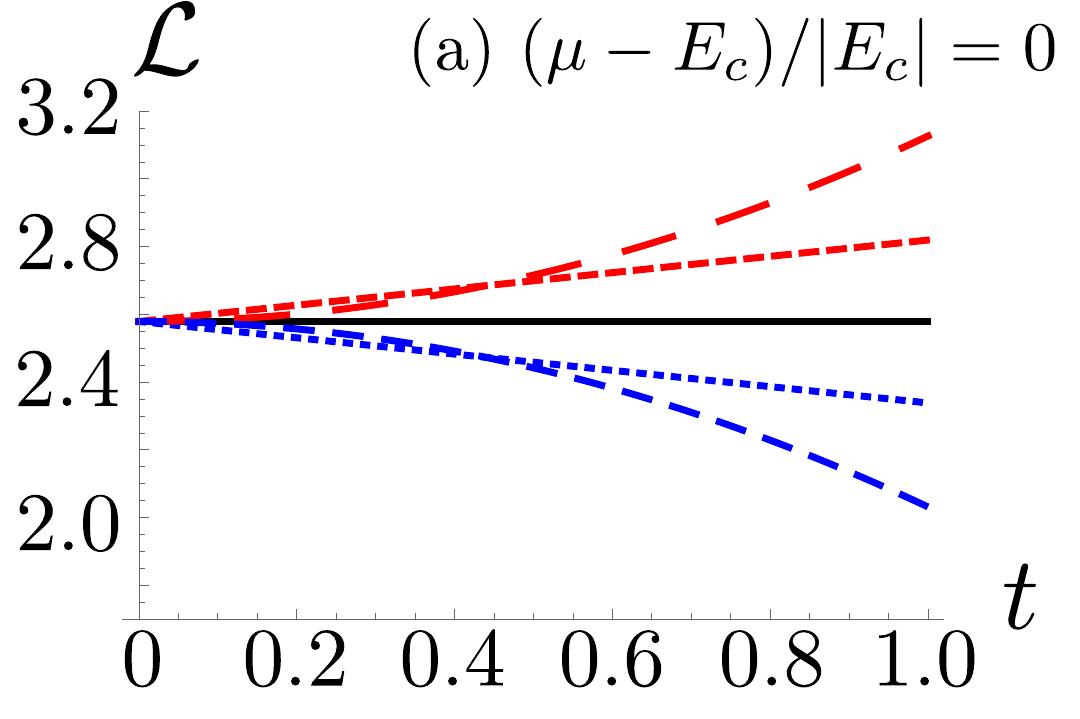}\\
    \vspace{\baselineskip}
    \includegraphics[width=5cm]{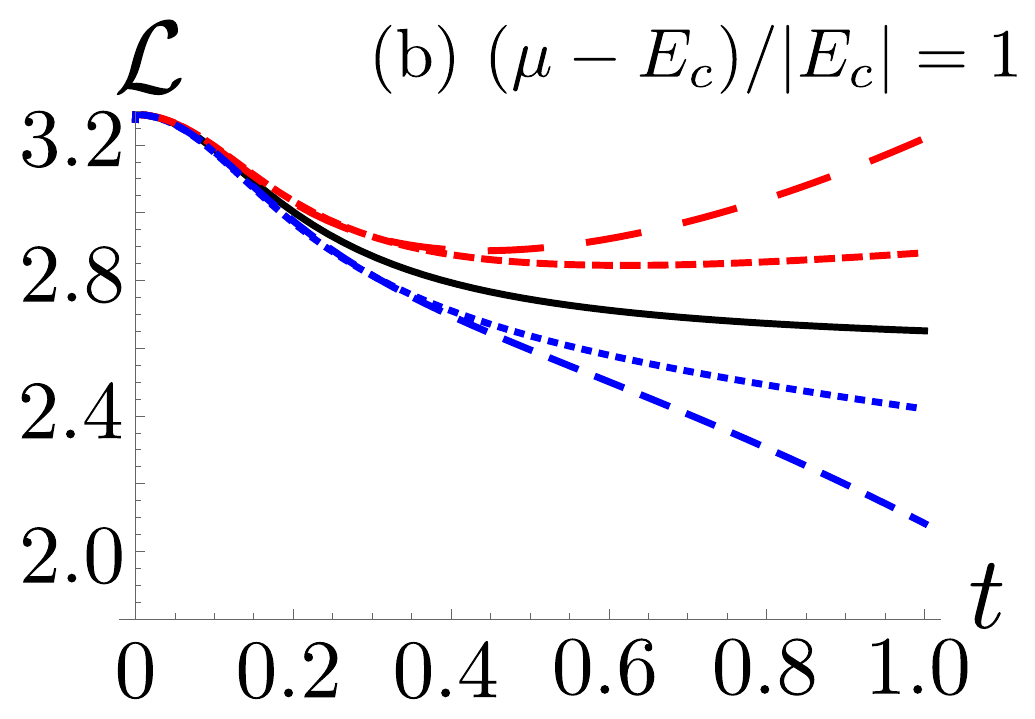}\\
    \vspace{\baselineskip}
    \includegraphics[width=5cm]{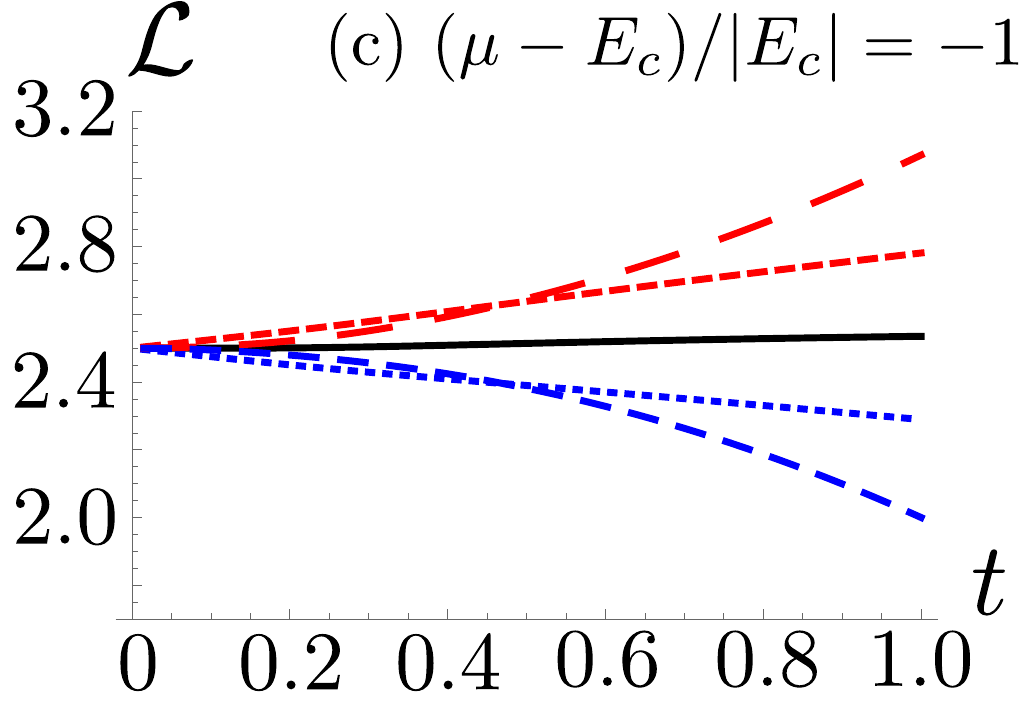}
    \caption{The Wiedemann-Franz ratio ${\cal L}$  in Eq.~\eqref{LLLL} versus $t = \kB T/|E_c|$ for $x=1.5$, $y=1$, with $a=0$ [solid  (black) curve], $a=0.05$ [dashed (red) curve] and $a=-0.05$ [dotted (blue) curve], as well as for $y=2$, with $a=0.01$ [largest-dashed (red) curve] and $a=-0.01$ [medium-dashed (blue) curve]. (a) $(\mu-E_c)/|E_c|=0$, (b) $(\mu-E_c)/|E_c|=1$, (c) $(\mu-E_c)/|E_c|=-1$.}
    \label{L0}
 \end{figure}

Figure \ref{L0} shows the temperature dependence of ${\cal L}$ with $a\ne 0$. Interestingly, the corrections introduce a temperature dependence of the Wiedemann-Franz ratio  even in regions where it was temperature independent in their absence.

\section{Finite-size effects}
\label{FSC}

Given Eqs.~(\ref{FSS}) and (\ref{FSS1}),  effects due to the finite size of the system can be taken into account upon replacing Eq.~(\ref{KK}) by
\begin{align}
K_n(x,z)&\rightarrow \widetilde{K}_n(
x,z,L)=\epsilon_L^x\int_{-z-\epsilon^{}_L}^{-z+\epsilon^{}_L} d\epsilon\epsilon^n\frac{1}{4\cosh^2(\epsilon/2)}\nonumber\\
&+\int_{-z+\epsilon^{}_L}^\infty d\epsilon\epsilon^n(\epsilon+z)^x\frac{1}{4\cosh^2(\epsilon/2)}\nonumber\\
&=K_n(x,z)+\epsilon_L^x\int_{-z-\epsilon^{}_L}^{-z+\epsilon^{}_L} d\epsilon\epsilon^n\frac{1}{4\cosh^2(\epsilon/2)}\nonumber\\
&-\int_{-z}^{-z+\epsilon^{}_L} d\epsilon\epsilon^n(\epsilon+z)^x\frac{1}{4\cosh^2(\epsilon/2)},
\label{KKL}
\end{align}
where the exponentially small  term in $\sigma^{}_0$  of  the insulating phase, which arises  from finite-size effects, is neglected.  Here
\begin{align}
\epsilon^{}_L=\frac{E_c}{\kB T}\left(\frac{\xi_0}{L}\right)^{1/\nu}
\label{epsl}
\end{align}
is a dimensionless energy which is a measure of the finite-size effects. This energy determines the width of the plateaus in Fig.~\ref{Finitesigma}.
Ignoring the corrections to scaling, we obtain the Seebeck coefficient, the figure of merit, and the Wiedemann-Franz ratio [Eqs.~(\ref{LLLL}),  (\ref{plotS}), and (\ref{plotZT})] as
\begin{align}
S=\frac{\kB}{|e|}\frac{{\widetilde K}_1(x,z,L)}{{\widetilde K}_0(x,z,L)},
\label{Ssize}
\end{align}
\begin{align}
ZT=\frac{{\widetilde K}_1(x,z,L)^2}{{\widetilde K}_0(x,z,L){\widetilde K}_2(x,z,L)-{\widetilde K}_1(x,z,L)^2}
\label{ZTsize}
\end{align}
and
\begin{align}
{\cal L}&=\left(\frac{\kB}{e}\right)^2\frac{{\widetilde K}_0(x,z,L){\widetilde K}_2(x,z,L)-{\widetilde K}_1(x,z,L)^2}{{\widetilde K}_0(x,z,L)^2}.
\label{Lsize}
\end{align}

At $z=0$, the first term in Eq.~(\ref{KKL}) vanishes for $n=1$ because the integrand is odd in $\epsilon$, so that
\begin{align}
&\widetilde{K}_1(x,0,L)-K_1(x,0)=\nonumber\\
&-\int_{0}^{\epsilon^{}_L} d\epsilon\epsilon^{1+x}(\epsilon+z)^x\frac{1}{4\cosh^2(\epsilon/2)},
\label{KK1}
\end{align}
becoming more negative as $\epsilon^{}_L$ increases.
In contrast, for $n=0, 2$ one has
\begin{align}
&\widetilde{K}_n(x,0,L)-K_n(x,0)=\nonumber\\
&\int_{0}^{\epsilon^{}_L} d\epsilon(2\epsilon^x_L-\epsilon^x)\epsilon^n(\epsilon+z)^x\frac{1}{4\cosh^2(\epsilon/2)},
\label{KK02}
\end{align}
becoming more positive as $\epsilon^{}_L$ increases. Since $\widetilde{K}_1$ decreases and $\widetilde{K}_0$ increases with $\epsilon^{}_L$, the Seebeck coefficient $S$ decreases with decreasing system size. Since both $\widetilde{K}_0$ and $\widetilde{K}_2$ increase, while $\widetilde{K}_1$ decreases, the electronic heat conductivity $\kappa$  (proportional to $\widetilde{K}_0\widetilde{K}_2-\widetilde{K}_1^2$) also increases with decreasing size, and therefore the figure of merit $ZT$ decreases with decreasing $L$. The situation is more complicated for the Wiedemann-Franz ratio ${\cal L}$: both numerator and denominator in Eq.~(\ref{Lsize}) increase with $\epsilon^{}_L$, and the result for ${\cal L}$ is not monotonic.

For $z\gg 1$ the finite-size corrections are very small: The difference
 \begin{align}
 &\widetilde{K}_n(x,z,L)-K_n(x,z)&\nonumber\\
 &\approx\epsilon_L^x\int_{-z-\epsilon^{}_L}^{-z+\epsilon^{}_L} d\epsilon\epsilon^n e^\epsilon%\nonumber\\
+\int_{-z}^{-z+\epsilon^{}_L} d\epsilon\epsilon^n(\epsilon+z)^x e^\epsilon
\label{finsiz}
\end{align}
is very small, of order $e^{-z}$, compared to the much larger expressions in Eq.~(\ref{A2}).
In contrast, for $z\ll -1$ both $K_n(x,z)$ [Eq.~(\ref{A11})] and the differences (\ref{finsiz}) for each $n=1,2,3,\cdots$ are exponentially small, of order $e^{-|z|}$, and therefore one expects significant finite-size corrections.
Using the same approximations as in Eq.~(\ref{A11}), we find
\begin{widetext}
\begin{align}
\widetilde{K}_0(x,z,L)  &= e^z \left[\epsilon_L^{x}(e^{\epsilon^{}_L}-e^{-\epsilon^{}_L}) + \Gamma(x+1,\epsilon^{}_L)\right]\equiv K_{00}, \nonumber\\
\widetilde{K}_1(x,z,L) &= e^z  \left\{ \epsilon_L^x [(1-\epsilon^{}_L-z)e^{\epsilon^{}_L} -(1-z)e^{-\epsilon^{}_L} ]+(x+1-z)\Gamma(x+1,\epsilon^{}_L) \right\}\equiv K_{10}-K_{00}z, \nonumber\\
\widetilde{K}_2(x,z,L) &=e^{z} \Big\{\epsilon_L^{x}\big[\big(\epsilon_L^2-2\epsilon^{}_L+2+2(\epsilon^{}_L-1)z+z^2\big)e^{\epsilon^{}_L}+(x\epsilon^{}_L -2+2z-z^2)e^{-\epsilon^{}_L}\big]
\nonumber\\
&+[(x+1)(x+2)-2(x+1)z+z^2]\Gamma(x+1,\epsilon^{}_L)\Big\}\equiv K_{20}-2K_{10}z+K_{00}z^2,
\label{Knegz}
\end{align}
\end{widetext}
where
\begin{align}
\Gamma(s,r) \equiv \int_r^{\infty}dt \  t^{s-1}e^{-t}
\end{align}
is the upper incomplete Gamma function. Note that $\Gamma(s+1,r)=s\Gamma(s,r)+r^s e^{-r}$.

\begin{figure}
    \includegraphics[width=6cm]{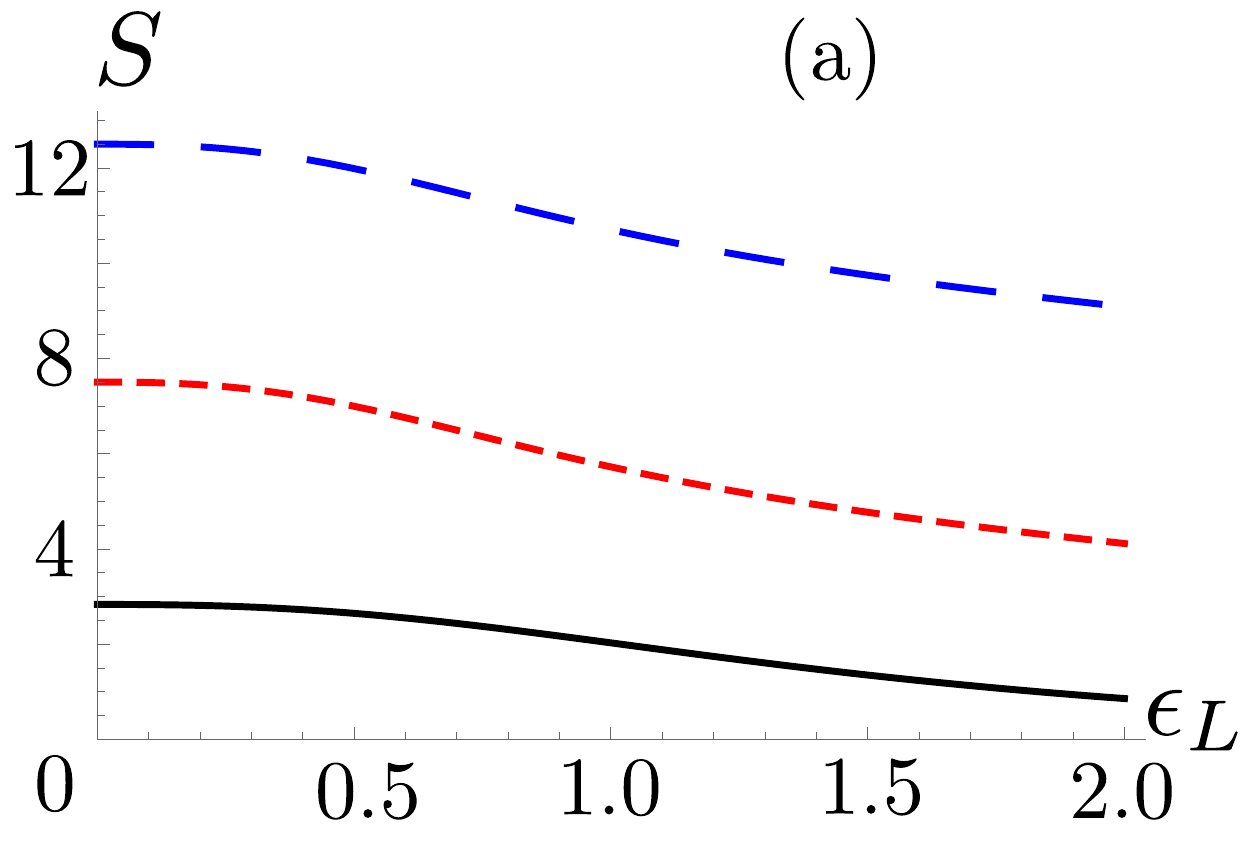}\\
    \vspace{\baselineskip}
    \includegraphics[width=6cm]{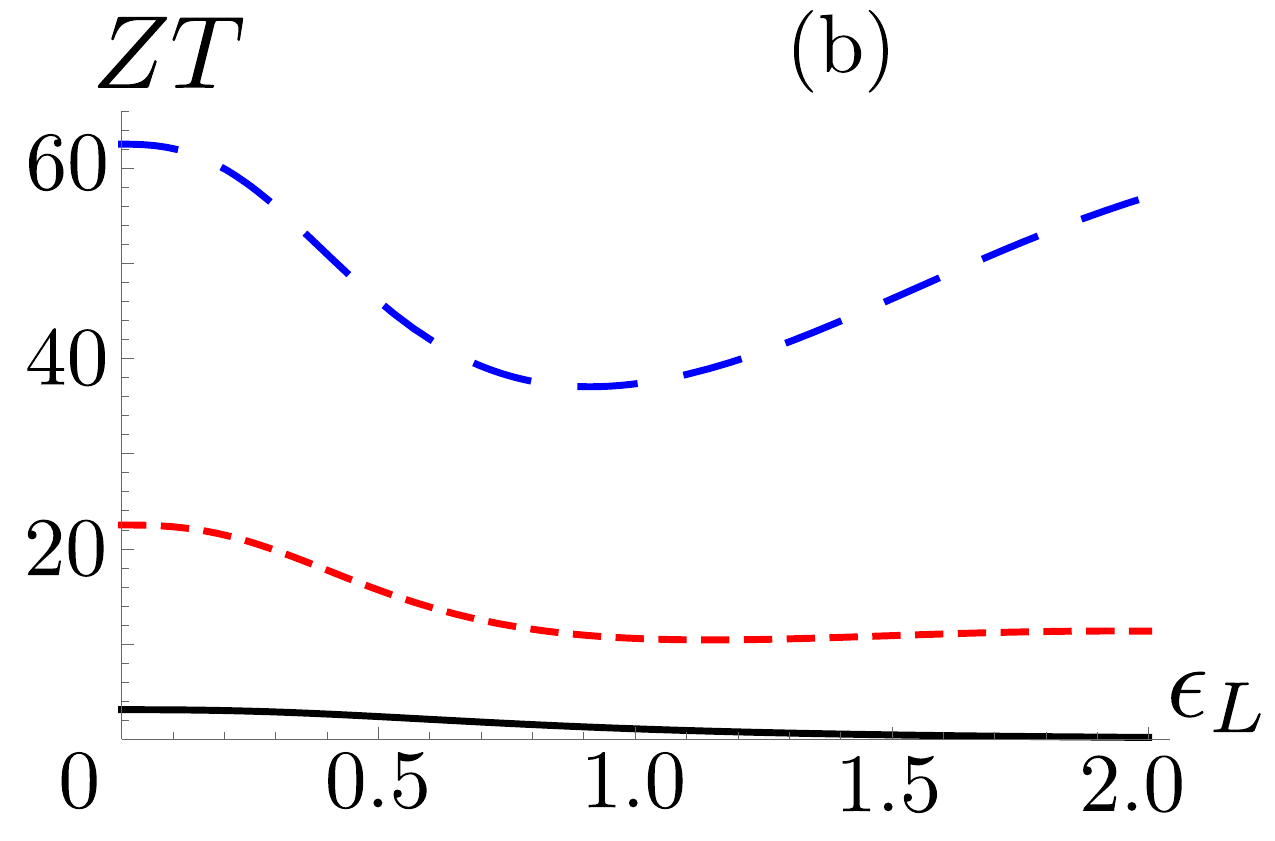}\\
    \vspace{\baselineskip}
     \includegraphics[width=6cm]{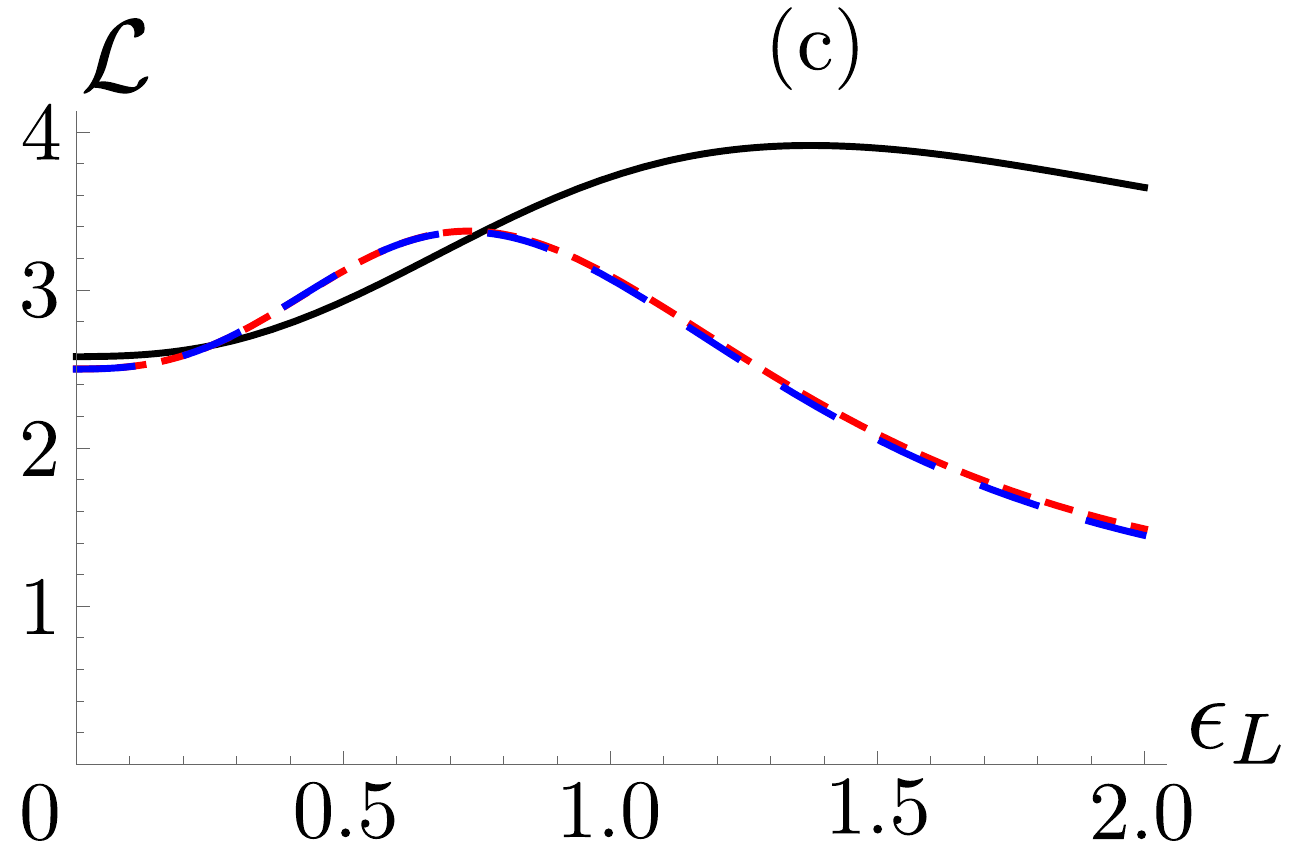}
    \caption{The Seebeck coefficient $S$ [in units of $\kB/|e|$, panel (a)], the figure of merit $ZT$ [panel (b)] and the Wiedemann-Franz ratio ${\cal L}$ [in units of $(\kB/|e|)^2$, panel (c)] as functions of the dimensionless energy  $\epsilon^{}_L$,   Eqs.~\eqref{Ssize}, \eqref{ZTsize}, and \eqref{Lsize},  for $x=1.5$  and $z=0$   [solid (black) curve], $z=-5$ [small-dashed (red) curve], and $z=-10$  [large-dashed (blue) curve].  Note that ${\cal L}$ is practically independent of $z$ for $z\ll -1$.
    }
    \label{Sep}
\end{figure}

Figure \ref{Sep} displays the size dependence of $S$, $ZT$, and ${\cal L}$, for several nonpositive values of $z$. We show the results only for small values of $\epsilon_L$, because the power laws are applicable only near the transition point.
As Fig.~\ref{Sep}(a) shows, the Seebeck coefficient $S$ continues to decrease with increasing $\epsilon^{}_L$ also for $z\ll -1$. An explicit calculation of Eqs.~(\ref{Knegz}) shows that $\widetilde{K}_0,~\widetilde{K}_1,~\widetilde{K}_2$, and $\kappa$ all increase with $\epsilon^{}_L$. However, as seen in Figs.~\ref{Sep}(b) and \ref{Sep}(c), their ratios for $ZT$ and for ${\cal L}$ are not monotonic in $\epsilon^{}_L$.
 In particular, at $z=-10$ the figure of merit becomes comparable to its value for the infinite system around $\epsilon^{}_L\approx 2$. For relatively large $\epsilon^{}_L$ ($1\leq \epsilon^{}_L\leq 2$) the figure of merit increases [Fig.~\ref{Sep}(b)] and the Wiedemann-Franz ratio decreases [Fig.~\ref{Sep}(c)] with decreasing size.\cite{power}

\begin{figure}
  \begin{center}
    \includegraphics[width=6cm]{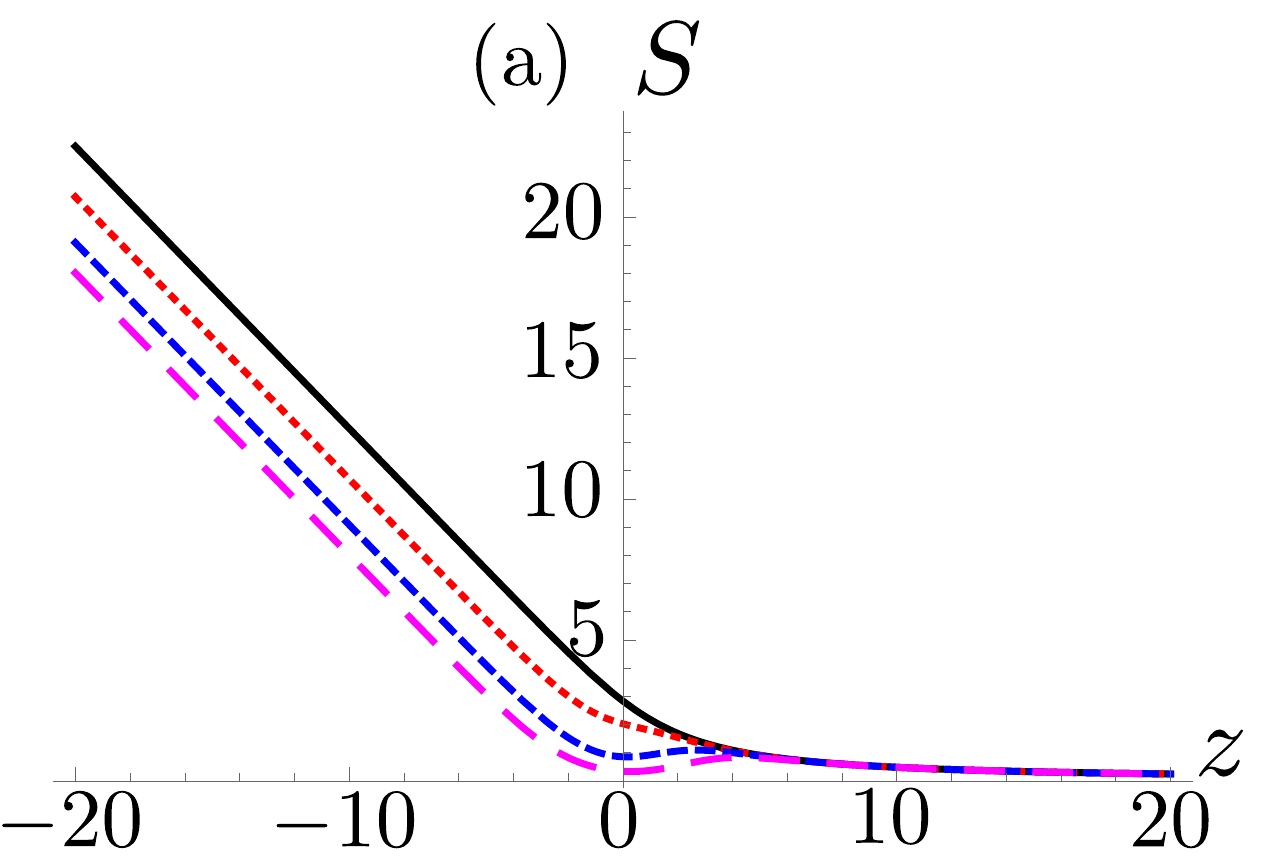}\\
    \vspace{\baselineskip}
    \includegraphics[width=6cm]{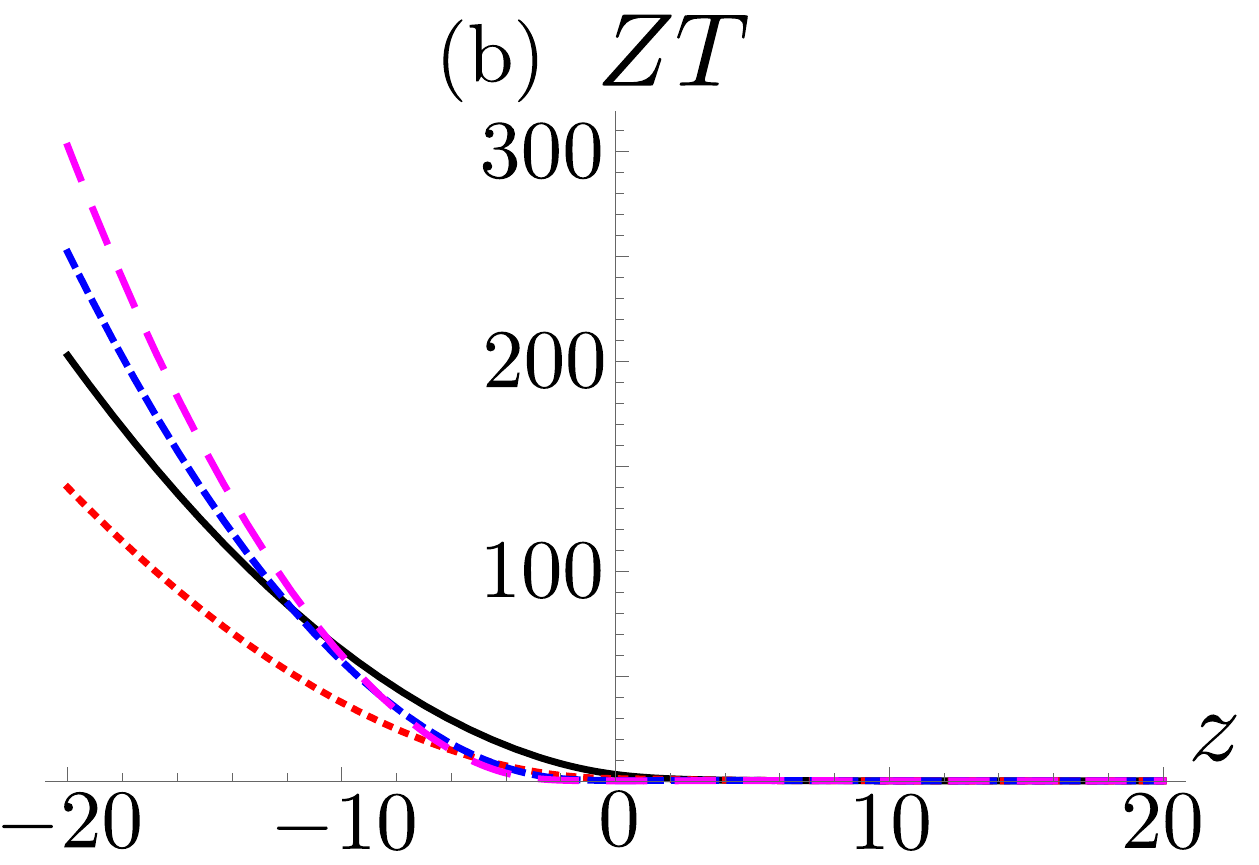}\\
    \vspace{\baselineskip}
    \includegraphics[width=6cm]{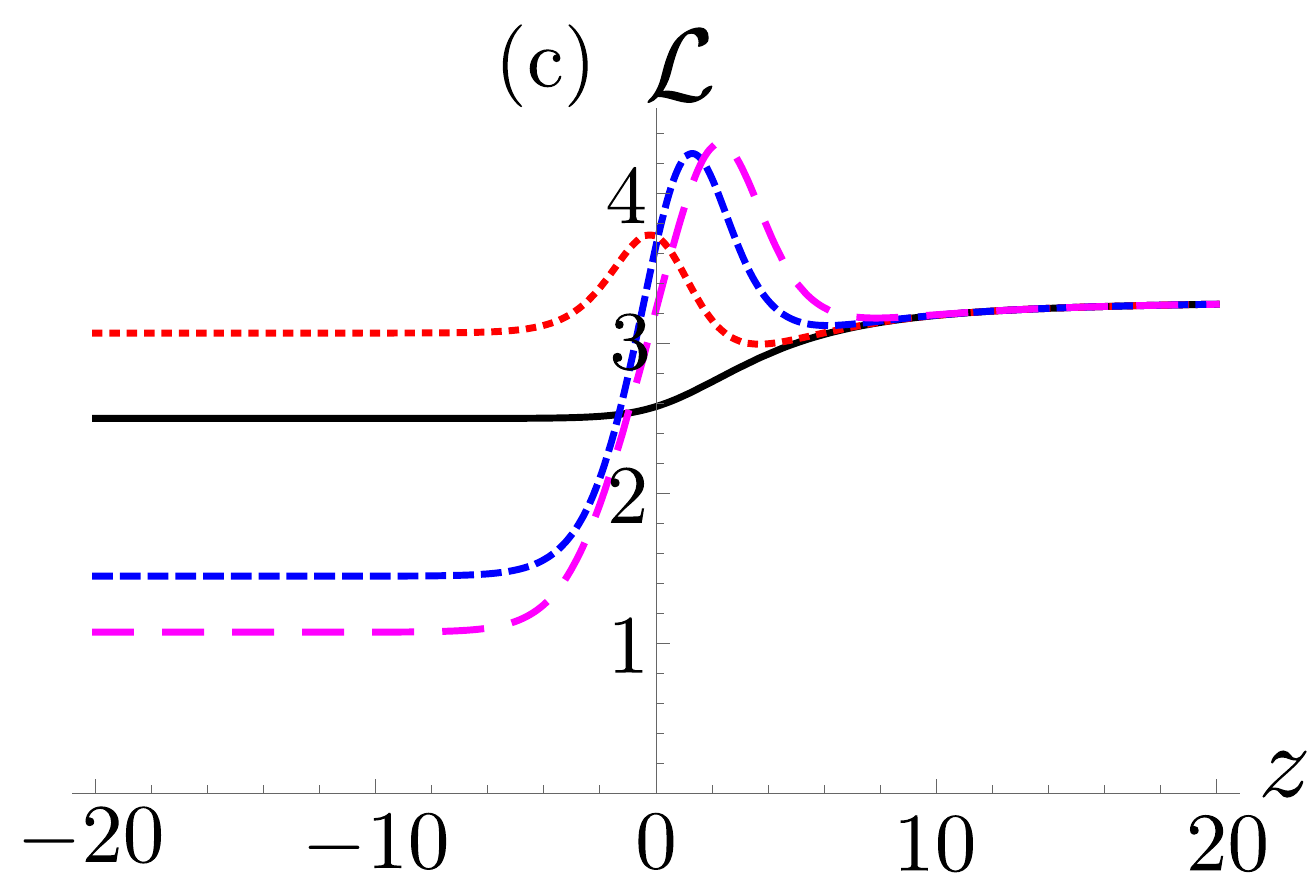}\\
    \caption{The Seebeck coefficient $S$ [in units of $\kB/|e|$, panel (a)], the figure of merit $ZT$ [panel (b)] and the Wiedemann-Franz ratio ${\cal L}$ [in units of $(\kB/|e|)^2$, panel (c)] versus $z=(\mu-E_c)/(\kB T)$ with $x=1.5$ for  $\epsilon_L=0$  [solid (black) curve], $\epsilon_L=1$ [dotted (red) curve],  $\epsilon_L=2$ [dashed (blue) curve], and  $\epsilon_L=3$ [large-dashed (magenta) curve]. }
    \label{SL}
  \end{center}
\end{figure}

Figure \ref{SL} displays $S$, $ZT$, and ${\cal L}$ versus $z$ for several   values of the system size.    As expected, for $z\gg 1$ all three quantities are almost size independent. For $z\ll -1$, Eqs.~(\ref{Knegz}) give
\begin{widetext}
\begin{align}
S &=\frac{L_{12}}{TL_{11}}= \frac{\kB}{|e|} \left\{\frac{\epsilon_L^x[(1-\epsilon^{}_L)e^{\epsilon^{}_L} -e^{-\epsilon^{}_L}]+(x+1)\Gamma(x+1,\epsilon^{}_L)}{\epsilon_L^{x}(e^{\epsilon^{}_L}-e^{-\epsilon^{}_L})
+\Gamma(x+1,\epsilon^{}_L)} -z\right\}.
\end{align}
\end{widetext}
The Seebeck coefficient is linear in $z$, with a size-independent slope. As seen in Fig.~\ref{SL}(a), the magnitude of $S$ decreases with decreasing size.

In contrast to $S$, and as already seen in Fig.~\ref{Sep}(b), $ZT$ exhibits a nontrivial nonmonotonic size dependence for $z<0$.
For large negative $z$, Eqs.~(\ref{ZT}) and (\ref{Knegz}) show that
\begin{align}
ZT=\frac{(K_{10}-K_{00}z)^2}{K_{00}K_{20}-K_{10}^2}.
\end{align}
Numerically, this function is found to decrease with $\epsilon^{}_L$ for small $\epsilon^{}_L$, but then to increase for $\epsilon^{}_L\gtrsim 1$, thus explaining the nonmonotonic size dependence of $ZT$. In this range, smaller devices are more efficient.

%%%%%%%%%%%%%%%%%%%

Figure \ref{SL}(c) presents the $z$ dependence of the Wiedemann-Franz ratio for several system sizes.
Again, there is almost no size dependence for $z\gg 1$. For large negative $z$ the ratio ${\cal L}$ approaches the $z$-independent limit ${\cal L}\approx (K_{00}K_{20}-K_{10}^2)/K_{00}^2$. As can be seen in Fig.~\ref{SL}(c), this value is also not monotonic in $\epsilon^{}_L$.

\section{A band with two mobility edges}

\label{TM}

As discussed in Sec.~\ref{Intro}, the case of two mobility edges is described by inserting  Eq.~(\ref{band}) into Eqs.~(\ref{LL}), to obtain
\begin{align}
L_{11} &= At^x {\cal K}_0(x,{\bar E}_c,{\bar \mu}),\nonumber\\
L_{12}&= A\frac{\kB T}{|e|}t^x {\cal K}_1(x,{\bar E}_c,{\bar \mu}),\nonumber\\
L_{22}&=  A\left(\frac{\kB T}{e}\right)^2 t^x {\cal K}_2(x,{\bar E}_c,{\bar \mu}),
\label{intL1}
\end{align}
where
\begin{align}
{\bar E}_c=E_c/(\kB T),\ \ \ {\bar \mu}=\mu/(\kB T),
\end{align}
(note: the chemical potential $\mu$ is now measured relative to the center of the band, $E=(E_{c1}+E_{c2})/2=0$) and
\begin{align}
{\cal K}_n(x,{\bar E}_c,{\bar \mu})=\int_{-{\bar \mu}-{\bar E}_c}^{-{\bar \mu}+{\bar E}_c}d\epsilon \frac{({\bar E}_c-|\epsilon+{\bar \mu}|)^x\epsilon^n}{4\cosh(\epsilon/2)}.
\label{Kband}
\end{align}

\begin{figure}
    \includegraphics[width=6cm]{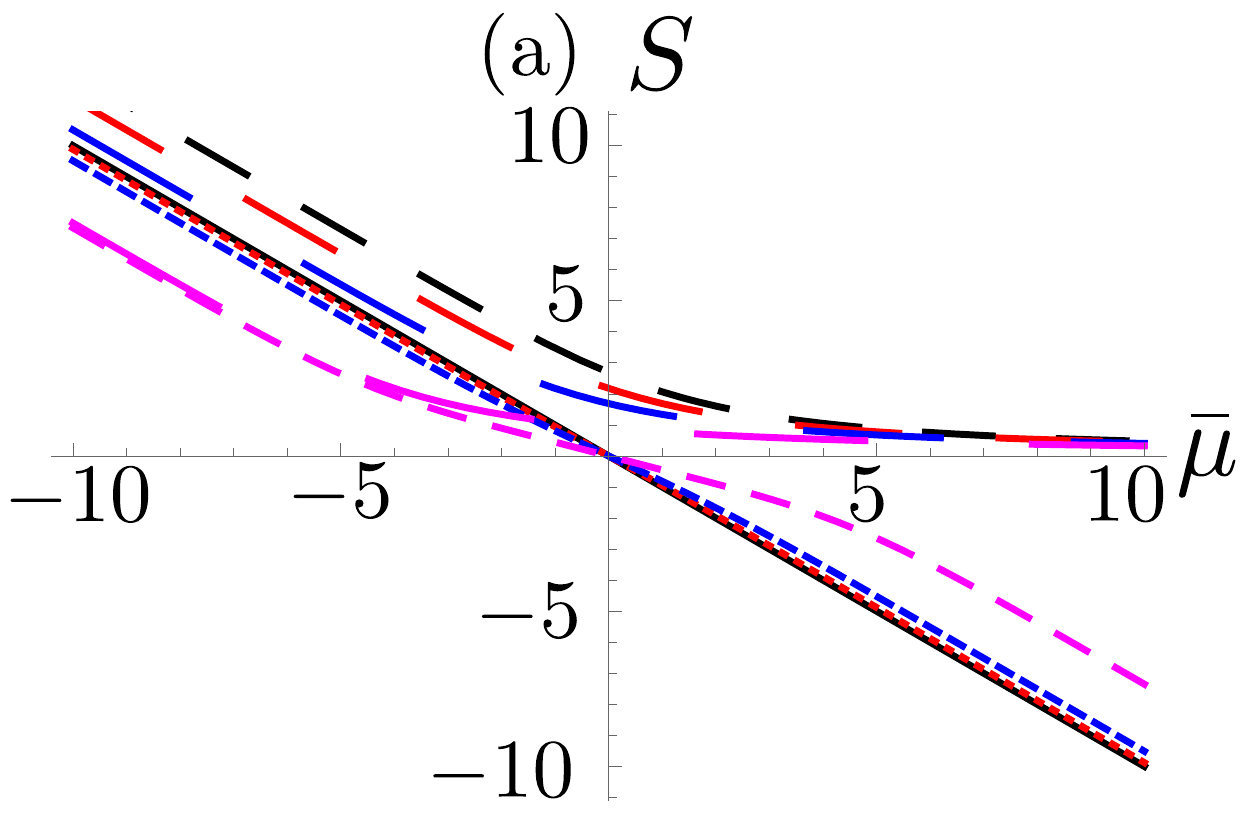}\\
     \vspace{\baselineskip}
     \includegraphics[width=6cm]{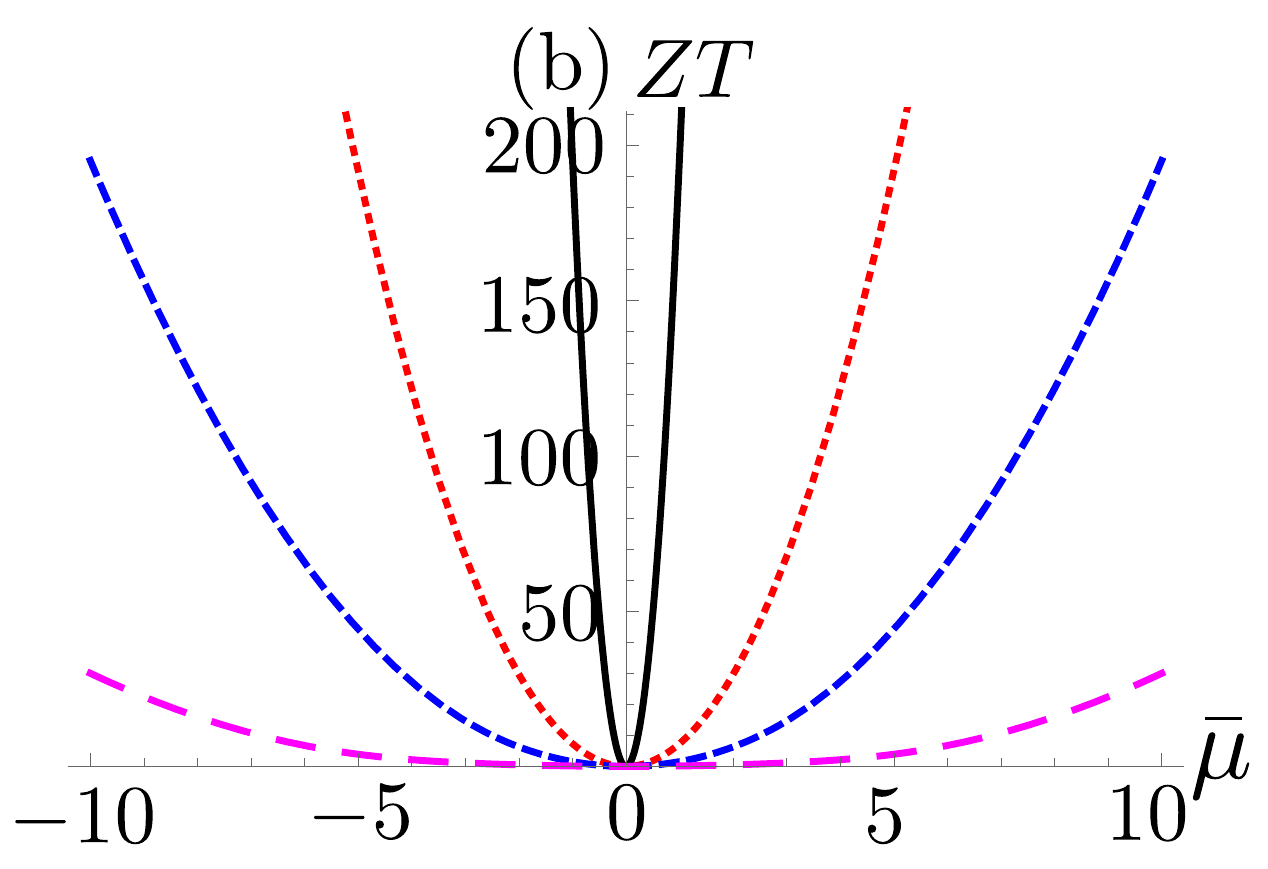}\\
      \vspace{\baselineskip}
       \includegraphics[width=6cm]{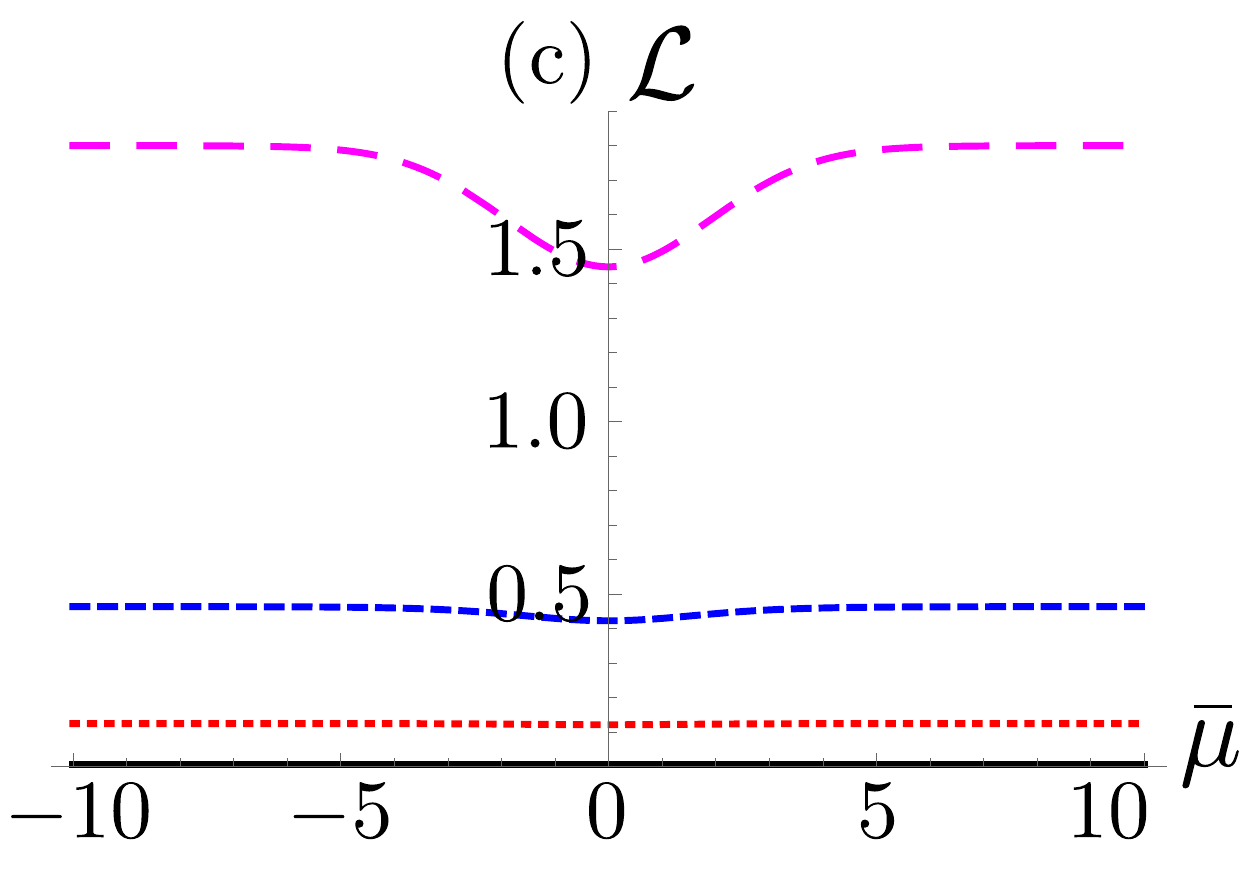}\\
    \caption{The Seebeck coefficient $S$ [in units of $\kB/|e|$, panel (a)], the figure of merit $ZT$ [panel (b)] and the Wiedemann-Franz ratio ${\cal L}$ [in units of $(\kB/|e|)^2$, panel (c)] versus ${\bar \mu}=\mu/(\kB T)$ for $x=1.5$ and for  ${\bar E}_c=E_c/(\kB T)=0.2$  [solid (black) line], $1$ [small-dashed (red) line], $2$ [medium-dashed (blue) line], and $5$ [large-dashed (magenta) line].
    The larger-dashed four curves in the plot of the Seebeck coefficient [panel (a)] are calculated with the single-threshold expression. }
    \label{SB}
\end{figure}

The Seebeck coefficient $S$, the figure of merit $ZT$, and the Wiedemann-Franz ratio ${\cal L}$ pertaining to this case, for specific values of $x$, ${\bar E}_c$, and ${\bar \mu}$, are displayed in Fig.~\ref{SB}. 
All  three plots exhibit a crossover around the mobility edges, $|{\bar \mu}| \sim \pm {\bar E}_c$. For $|{\bar \mu}| < {\bar E}_c$, $S$, and $ZT$ are small, while ${\cal L}$ has a minimum. For $|{\bar \mu}| > {\bar E}_c$, $|S|$, and $ZT$ increase with $|{\bar \mu}|$, while ${\cal L}$ approaches a plateau at values which decrease with decreasing ${\bar E}_c$. This latter decrease indicates  a decrease in the heat conductivity $\kappa$.  As expected by Mahan and Sofo,\cite{mahan}, this causes a fast increase in the figure of merit.\cite{power} 

The dashed lines in Fig.~\ref{SB}(a) show the Seebeck coefficient calculated with the single-threshold expression, Eq.~(\ref{eq:0Tcond}). As might be expected, the two calculations coincide for $\mu<0$ in the limit $E_c\gg \kB T$; the difference between Eq.~(\ref{Kband}) and the single-band case, Eq.~(\ref{KK}), becomes of order $e^{-|{\bar \mu}|}$, due to the decay of $1/[4\cosh^2(\epsilon/2)]\approx e^{-|\epsilon|}$.

 When  $|{\bar \mu}| \gg {\bar E}_c$,  Eq.~(\ref{Kband})  can be expanded  to the lowest order in ${\bar E}_c$, as in
\begin{align}
{\cal K}_n(x,{\bar E}_c,{\bar \mu})\approx {\bar E}_c^{x+1}(-{\bar \mu})^n/[2\cosh^2({\bar \mu}/2)].
\end{align}
With this approximation one indeed finds that the electronic heat conductivity vanishes, ${\cal K}_0{\cal K}_2-{\cal K}_1^2=0$, and therefore ${\cal L}=0$ and $ZT=\infty$. It should be kept in mind that the the phononic heat conductivity should be included in the definition of the figure of merit, in particular in this limit. \cite{power} In this regime $S=-(\kB/|e|){\bar \mu}$, in full agreement with Fig.~\ref{SB}.

\section{Conclusions and discussion}

\label{DIS}

For an isolated localization edge in an infinite system, we have presented a detailed analysis of the universal functions which describe   various thermoelectric coefficients as functions of the distance of the mobility edge energy, $E_c$, from the chemical potential, in units of $\kB T$, $z=(\mu-E_c)/(\kB T)$. In particular, we corrected certain expressions which appeared in the earlier literature and added several new approximants for specific values of $z=(\mu-E_c)/(\kB T)$. We also added a new detailed discussion on the ``insulating" region $z<0$, where we find large values of the Seebeck coefficient and the figure of merit. 

In addition, we have introduced corrections to scaling, which generate additional nonuniversal temperature dependences of the various coefficients. Such corrections may be needed when measurements move away from the  vicinity of the localization edge.

In the second part of the paper, we have introduced finite-size effects. These results are highly relevant to future mesoscopic thermoelectric devices.
As the system becomes smaller, we found that  the Seebeck coefficient becomes smaller, but  the figure of merit and the Wiedemann-Franz ratio behave nonmonotonically with the system size. In particular, electronic heat conductivity decreases and  the figure of merit increases for smaller systems. It would be very interesting to probe these predictions experimentally.

Surprisingly, we could not find earlier analyses of thermoelectricity in the common situation of a disorder-generated narrow zero-temperature mobility range near the center of an energy band. It turns out that disorder enhances the thermoelectric efficiency, particularly near the value of the disorder  at which the mobility range shrinks to zero and the system is always an insulator.

It should be emphasized that our whole paper considered only noninteracting electrons and ignored any  inelastic processes.   The issue of inelastic processes is particularly important. One way to include such processes is to replace the size $L$ by the inelastic length $L_i\sim T^{-p}$.\cite{sivan1986,kapitulnik1992}  We leave this discussion for the future.

Although we find a large electronic figure of merit for large negative $z$, i.e., deep in the insulating phase, this result should be considered with care. First, we ignored the phononic thermal conductivity, $\kappa^{}_{ph}$. Since this conductivity should be added to $\kappa$ in the denominator of Eq.~(\ref{ZT}), it becomes important when the electronic $\kappa$ becomes very small, which happens for large negative $z$. We expect a significant decrease of our calculated $ZT$ when $\kappa^{}_{ph}/\kappa \geq 1$.\cite{ora} Secondly, at finite temperatures and deep in the insulating regime the main mechanism for electronic conductivity is the phonon-assisted  (or variable-range) hopping.\cite{ora1,ying, suri,pichard}  Although the mechanism described in the present paper should dominate the charge and heat transport at sufficiently low temperatures, there must exist a cross-over temperature, above which the transport is dominated by the phonon-assisted  (or variable-range) hopping. This cross-over temperature depends on the system.  

Although noninteracting electronic states are fully localized in a random macroscopic two-dimensional electron gas, large localization lengths and sufficiently small systems have recently been observed to exhibit large thermoelectric and electric transport.\cite{pepper} It may be interesting to apply our analysis in the vicinity of such an effective '2D metal-insulator transition'.

\vspace{.5cm}
{\bf Acknowledgements}

We are grateful to Prof.~Tomi Ohtsuki for useful discussion. We also acknowledge the hospitality of Beijing Computational Science Research Center (CSRC), China, where part of this work was accomplished.
The work at Ben Gurion University was supported by the Israeli Science Foundation and by
the infrastructure program of Israel Ministry of Science and
Technology under Contract No. 3-11173.
K.Y.~is supported by the Advanced Leading Graduate Course for Photon Science (ALPS), the University of Tokyo as well as by a Grant-in-Aid for Japan Society for the Promotion of Science (JSPS) Fellows (Grant No. 16J11542).
N.H.~is supported by Kakenhi Grant No. 15K05200, No. 15K05207, and No. 26400409 from the Japan Society for the Promotion of Science.

\vspace{1cm}

\appendix

\begin{widetext}

\section{Approximations}\label{AppA}

In this Appendix we derive approximate analytic expressions for the integrals $K_n(x,z)$, defined in Eq.~(\ref{KK}), and for the Onsager coefficients. The results are used in Sec.~\ref{IS}  to obtain analytic approximants for  the Seebeck coefficient, the figure of merit, and the Wiedemann-Franz ratio.

\subsection{Low temperatures and $\mu- E_c\gg \kB T$}

 When $\kB T\ll E-\mu$, i.e., $\epsilon\gg 1$,  the symmetric function $1/\cosh^2(\epsilon)$ decays exponentially for $|\epsilon|>3$, and  the  main contributions to integrals of the form $\int_{-z}^\infty d\epsilon G(\epsilon)/[4\cosh^2(\epsilon)]$, for an arbitrary  function $G(\epsilon)$ which does not change rapidly around $\epsilon=0$, come from small $\epsilon$.  On the other hand, the lower limit of the integral is large and negative, because $z=(\mu-E_c)/(\kB T)\gg 1$. Therefore we can replace the lower limit of the integrals by $-\infty$ and expand $G(\epsilon)$ in powers of $\epsilon$. This procedure is equivalent to the Sommerfeld expansion. \cite{ashcroft}
The  result is
\begin{align}
&\int_{-z}^{\infty}d\epsilon G(\epsilon)/[4\cosh^2(\epsilon/2)]
=G(0)+\frac{\pi^2}{6} G^{''}(0)+ \frac{7\pi^4}{360} G^{''''}(0) +\cdots.
\end{align}
 For $G(\epsilon)=\epsilon^n(\epsilon+z)^n$   one finds
\begin{align}
K_n(x,z)&=z^x\Big(\delta_{n,0}+\frac{\pi^2}{6}\big[2\delta_{n,2}+2\delta_{n,1}x z^{-1}+x(x-1)\delta_{n,0}z^{-2}\big]
+\frac{7\pi^4}{360}\big[24\delta_{n,4}+24\delta_{n,3}xz^{-1}+12\delta_{n,2}x(x-1)z^{-2}\nonumber\\
&+4\delta_{n,1}x(x-1)(x-2)z^{-3}
+\delta_{n,0}x(x-1)(x-2)(x-3)z^{-4}\big]\Big)+\cdots .
\label{A2}
\end{align}
 The linear-response coefficients introduced in Eqs.~(\ref{Eq1}),  (\ref{SSS}), and (\ref{LL})
 become series in $1/z^2$,
\begin{align}
L_{11} &=A(tz)^x\left[1 +at^y+ \frac{\pi^2}{6} x(x-1)\frac{1}{z^2}+{\cal O}[1/z^4,(at^y)^2]\right], \nonumber\\
L_{12} &= A\frac{\kB T}{|e|}\frac{\pi^2}{3}
 t^x z^{x-1}\left[x+a(x+y)t^y
 +\frac{7\pi^2}{30}x(x-1)(x-2) \frac{1}{z^2}+{\cal O}[1/z^4,(at^y)^2]\right], \nonumber\\
L_{22} &= A\left(\frac{\kB T}{e}\right)^2\frac{\pi^2}{3}(tz)^x\left[1+at^y+\frac{7\pi^2}{10} x(x-1)\frac{1}{z^2}+{\cal O}[1/z^4,(at^y)^2]\right],
 \label{largez}
\end{align}
where $t=\kB T/|E_c|$, $A$ is the coefficient of the leading term, and $a$ is the coefficient of the next term  in the zero-temperature conductivity, Eq.~(\ref{corr}).
 The Seebeck coefficient, the figure of merit, and the Wiedemann-Franz ratio are then given by
 \begin{align}
S&=\frac{L_{12}}{TL_{11}}=\frac{\kB}{|e|}\frac{\pi^2}{3}\frac{1}{z}\left[x
+\frac{\pi^2}{15}x(x-1)(x-7)\frac{1}{z^2}
+ayt^y+{\cal O}[1/z^4,(at^y)^2]\right],
\label{SlowT}
\end{align}
\begin{align}
ZT=\frac{L_{12}^2}{L_{11}L_{22}-L_{12}^2}=\frac{\pi^2x^2}{3z^2}\left[1-\frac{\pi^2}{15}(x^2+8x-14)\frac{1}{z^2}+2\frac{y}{x}at^y
+{\cal O}[1/z^4,(at^y)^2]\right],
\label{ZTlowT}
\end{align}
and
\begin{align}
 {\cal L}&=\frac{\kB^2}{e^2}\frac{\pi^2}{3}\left[1+\frac{\pi^2}{15}x(3x-8)\frac{1}{z^2}+{\cal O}[1/z^4,(at^y)^2]\right].
\label{LlowT}\end{align}

\subsection{ At the mobility edge:  $|z|=0$}

At $\mu=E_c$, i.e., $z=0$ the integrals in Eqs.~(\ref{KK}) can be calculated analytically. Expanding the integrand we find 
\begin{align}\label{Inu}
K_n(x,0)&=\int_0^\infty d\epsilon \epsilon^{x+n}\sum_{m=1}^\infty(-1)^m m e^{-m\epsilon}
=\sum_{m=1}^\infty (-1)^m m^{-(x+n)}\Gamma(x+n+1)\nonumber\\
&=\eta(x+n)(x+n)\Gamma(x+n)=(x+n)I_{x+n}
\end{align}
for all $n+x>-1$, where $\Gamma(u)$ is the Gamma function, $\eta(u)=(1-2^{1-u})\zeta(u)$ is the Dirichlet  eta function,
\begin{align}
I_u\equiv(1-2^{1-u})\zeta(u)\Gamma(u),
\label{IInu}
\end{align}
and $\zeta(u)$ is the Riemann zeta function.
At $u=1$ the zeta function diverges, but the expression for $I_u$  is continuous,  with $I_1=\ln 2$.
 At times (e.g., Ref.~\onlinecite{enderby1994}), the form  $I_u=\int_0^\infty dx x^{u-1}/(1+e^x)$   is used; however, this equality   is valid only for $u>0$, while Eq.~(\ref{IInu}) is valid for all $u>-1$.

\subsection{ High temperatures:  $|z|\ll 1$}

For $|z|\ll 1$ we expand $K_n(x,z)$ in powers of $z$. The leading-order term comes from
\begin{align}
\left. \frac{\partial K_n(x,z)}{\partial z}\right|_{z=0}=xK_{n-1}(x,0)=x(n+x-1)I_{n+x-1}.
\end{align}
Thus,
\begin{align}
K_n(x,z)&=K_n(x,0)+x K_{n-1}(x,0)z+{\cal O}(z^2)
=(n+x)I_{n+x}+x(n+x-1)I_{n+x-1}z+{\cal O}(z^2),
\end{align}
and the Onsager coefficients  [see Eqs.~(\ref{Eq1}) and (\ref{LL})] become
\begin{align}
L_{11}&=At^x \Big\{xI_x+x(x-1)I_{x-1}z
+at^y\big[(x+y)I_{x+y}+(x+y)(x+y-1)I_{x+y-1}z\big]+{\cal O}[z^2,(at^y)^2]\Big\},\nonumber\\
L_{12}&=A\frac{\kB T}{|e|}t^x \Big\{(x+1)I_{x+1}+x^2I_xz
+at^y\big[(x+y+1)I_{x+y+1}+(x+y)^2I_{x+y}z\big]+{\cal O}[z^2,(at^y)^2]\Big\},\nonumber\\
L_{22}&=A\left(\frac{\kB T}{e}\right)^2t^x\Big\{(x+2)I_{x+2}+x(x+1)I_{x+1}z
+at^y\big[(x+y+2)I_{x+y+2}
+(x+y)(x+y+1)I_{x+y+1}z\big]\nonumber\\
&+{\cal O}[z^2,(at^y)^2]\Big\}.
\label{LhighT}
\end{align}
In this regime, the leading corrections to  the Seebeck coefficient $S$ are
\begin{align}
S&\approx\frac{\kB}{|e|}\frac{(x+1)I_{x+1}+x^2I_xz+at^y(x+y+1)I_{x+y+1}}{xI_x+
x(x-1)I_{x-1}z+at^y(x+y)I_{x+y}}=S_0-S_1z+S_2 at^y, \label{ShighT1}
\end{align}
where
\begin{align}
 &S_0=\frac{\kB}{|e|}\frac{(x+1)I_{x+1}}{xI_x},\ \ \
 S_1=-\frac{\kB}{|e|}\left[x-\frac{(x^2-1)I_{x-1}I_{x+1}}{xI_x^2}\right],\nonumber\\
&S_2=\frac{\kB}{|e|}\left[\frac{(x+y+1)I_{x+y+1}}{xI_x}-\frac{(x+y)(x+1)I_{x+y}I_{x+1}}{x^2I_x^2}\right],
 \label{ShighT}
\end{align}
with further corrections of order $z^2$, $zat^y$, and $(at^y)^2$, and those to the figure of merit $ZT$ are
\begin{align}
&ZT \approx Z_0+Z_1z+Z_2at^y, \label{ZThighT1}
\end{align}
where
\begin{align}
&Z_1=Z_0\Big[\frac{2 x^2 I_x}{(x + 1) I_{x + 1}}
- \frac{x (x - 1) (x + 2) I_{x -1} I_{x + 2} -
    x^2 (x + 1) I_x I_{x + 1}}{x (x + 2) I_x I_{x + 2} - (x + 1)^2 I_{x + 1}^2}\Big],\nonumber\\
&Z_2=Z_0\Big\{ \frac{[(x+2)xI_{x+2}I_{x}-(x+1)^2I_{x+1}^2][2(x+y+1)(x+1)I_{x+y+1}I_{x+1}]}{[(x+2)xI_{x+2}I_{x}-(x+1)^2I_{x+1}^2](x+1)^2I_{x+1}^2} \notag \\
&- \frac{[(x+y)(x+2)I_{x+y}I_{x+2}+(x+y+2)xI_{x+y+2}I_x -2(x+y+1)(x+1)I_{x+y+1}I_{x+1}] }{[(x+2)xI_{x+2}I_{x}-(x+1)^2I_{x+1}^2]}\Big\},
 \label{ZThighT}
\end{align}
with $Z_0$ given in Eq.~(\ref{eq:ZTz0}).
 The Wiedemann-franz ratio is expanded as
\begin{align}
{\cal L}\approx{\cal L}_{00}(x)-{\cal L}_1(x)z+{\cal L}_2at^y,
\label{WLhighT}
\end{align}
where\cite{villagonzalo1999}
\begin{align}
{\cal L}_{00}(x)=\Big(\frac{\kB}{e}\Big)^2\Big[\frac{(x+2)I_{x+2}}{xI_x}-\frac{(x+1)^2I_{x+1}^2}{(xI_x)^2}\Big],
\label{LhighT1}
\end{align}
while
\begin{align}
{\cal L}_1(x)=2S_0S_1/Z_0+(S_0/Z_0)^2Z_1,\ \ \ {\cal L}_2(x)=2S_0S_2/Z_0-(S_0/Z_0)^2Z_2.
\end{align}
Surprisingly, in the range $0<x<2$, ${\cal L}_{00}(x)$ is very close to the linear approximant $(\kB/e)^2(1.38+0.81x)$.

\subsection{Low temperatures and $\mu-E_c\ll -\kB T$}

 When $z=(\mu-E_c)/(\kB T)$ is  large and negative,  the lower bound of $K_n$ in Eq.~(\ref{KK}) is large and positive, and therefore the integration variable $\epsilon$ is always very large, $\epsilon\gg 1$. We can then use the expansion
\begin{align}
\frac{1}{4\cosh(\epsilon/2)}=e^{-\epsilon}-2e^{-2\epsilon}+{\cal O}(e^{-3\epsilon}).
\end{align}
Writing
\begin{align}
K_n(x,z)=\int_0^\infty du(u-z)^nu^x(e^{z-u}-2e^{2z-2u}+\cdots),
\end{align}
and using $\int_0^\infty du u^xe^{-mu}=\Gamma(1+x)/m^{1+x}$, we find
\begin{align}
K_0(x,z)&=e^z\Gamma(1+x)[1-e^{z}/2^x+
{\cal O}(e^{2z})],\nonumber\\
K_1(x,z)&=e^z\Gamma(1+x)\big[1+x-z-e^{z}(1+x-2z)/2^{1+x}+
{\cal O}(e^{2z})\big],\nonumber\\
K_2(x,z)&=e^z\Gamma(1+x)\big\{(x+1)(x+2)-2(x+1)z+z^2
-e^z[(x+1)(x+2)-4(x+1)z+4z^2]/2^{x+2}]+
{\cal O}(e^{2z})\big\}.
\label{A11}
\end{align}

With these approximations, we arrive at
\begin{align}
L_{11}&=At^x e^z\Big[\Gamma(1+x)(1-e^z/2^{x})
+at^y\Gamma(1+x+y)(1-e^z/2^{x+y})+{\cal O}(e^{2z},a^2t^{2y})\Big],\nonumber\\
L_{12}&=A\frac{\kB T}{|e|}t^x e^z \Big[\Gamma(1+x)\big(1+x-z-e^z(1+x-2z)/2^{1+x}\big)
\nonumber\\&+at^y\Gamma(1+x+y)(1+x+y-z-e^z(1+x+y-2z)/2^{1+x+y}+{\cal O}(e^{2z},a^2t^{2y})\Big],\nonumber\\
L_{22}&=A\left(\frac{\kB T}{e}\right)^2t^x e^z \Big\{\Gamma(1+x)[(1+x)(2+x)-2(x+1)z+z^2
-e^z[(x+1)(x+2)\nonumber\\
&-4(x+1)z+4z^2]/2^{x+2}+at^y\Gamma(1+x+y)[(1+x+y)(2+x+y)\nonumber\\
&-2(1+x+y)z+z^2-
e^z[(1+x+y)(2+x+y)-4(1+x+y)z+4z^2]/2^{x+y+2}+{\cal O}(e^{2z},a^2t^{2y})\Big\},
\label{Lnegz}
\end{align}
and thus
\begin{align}
S=\frac{\kB}{|e|}\left[(1+x-z)+ \frac{1+x}{2^{x+1}}e^z+ayt^y\frac{\Gamma(1+x+y)}{\Gamma(1+x)}+{\cal O}(e^{2z},a^2t^{2y})\right]
\label{Snegz}
\end{align}
and
\begin{align}
&ZT = \frac{(1+x-z)^2}{1+x}+ \frac{(1+x-z)[x(3+x-z)+2(1+z)]}{2^{2+x}(1+x)}e^{z}\nonumber\\
&+at^y\frac{(1+x-z)\Gamma(1+x+y)}{(1+x)^2\Gamma(1+x)}\Big[4+4x^2+x(8+3y+y^2-4z)-y(-3+z)-y^2(-1+z)-4z\Big]+\mathcal{O}(e^{2z}, a^2t^{2y}). \label{ZTnegz}
\end{align}
The expression for ${\cal L}$ is given in Eq.~(\ref{LLnegz}).

\end{widetext}

%%%%%%%%%%%%%%%%%%%%%%%%%


\begin{thebibliography}{99}
\bibitem{com} Many authors define $S$ with a minus sign for electrons and $S>0$ for holes.  We define it with the opposite sign for the simplicity of presentation.
\bibitem{com1} This function should vary smoothly near $E=0$, but such a change has only a small effect on the results.
\bibitem{power} Note, however, that the output power from the thermoelectric device, $j_eV$, vanishes in the adiabatic limit in which the figure of merit diverges to infinity and the efficiency reaches the Carnot value. This is known as  power-efficiency trade-off.\cite{benenti-rev,brandner1, brandner2, shiraishi1, shiraishi2, pietzonka} Therefore, increasing disorder increases the (maximal) efficiency while decreasing the power, and one should consider this trade-off when planning experiments.  %In particular, for the tight coupling limit, in which the figure of merit diverges and the maximal efficiency becomes the Carnot one, the power becomes zero, which is useless. }
\end{thebibliography}
\end{document}